%% file: mumug.tex
\documentclass[smallextended]{svjour3}
\usepackage{amsmath}
\usepackage{amssymb}
\usepackage{slashbox}
\usepackage{url}
\usepackage{hyperref}
\begin{document}

\input FEYNMAN

\title{MUMUG: a fast Monte Carlo generator for radiative muon pair 
production (a pedagogical tutorial)}
\titlerunning{MUMUG: a fast Monte Carlo generator for the process
$e^+e^-\to\mu^+\mu^-\gamma$}
\author{Z.~K.~Silagadze}
\institute{Budker Institute of Nuclear Physics and Novosibirsk State 
University, Novosibirsk 630090, Russia
\email{Z.K.Silagadze@inp.nsk.su} }

\date{Received: date / Accepted: date}

\maketitle

\begin{abstract}
A fast leading-order Monte Carlo generator for the process $e^+e^-\to\mu^+
\mu^-\gamma$ is described. In fact, using the $e^+e^-\to\mu^+\mu^-\gamma $ 
process as an example, we provide a pedagogical demonstration of how 
a Monte Carlo generator can be created from scratch. The $e^+ e^- \to 
\mu^+ \mu^- \gamma$ process was chosen, since in this case we are not faced 
with either too trivial or too difficult a task. Matrix elements are 
calculated using the helicity amplitude method. Monte Carlo algorithm uses 
the acceptance-rejection method with an appropriately chosen simplified 
distribution that can be generated using an efficient algorithm. We provide 
a detailed pedagogical exposition of both the helicity amplitude method and 
the Monte Carlo technique, which we hope will be useful for high energy 
physics students.
\end{abstract}

\section{introduction}
Budker Institute of Nuclear Physics (Novosibirsk) is currently developing an 
expensive and long-term (~10 years) project for a super charm-tau factory  
which will require many new accelerator technologies. To test and study these 
technologies at the first stage of the super charm-tau factory project, it is 
planned to build an inexpensive, low-energy machine called $\mu \mu$-tron 
\cite{1}. It is so named because, in addition to purely accelerator studies, 
it allows solving the interesting physical problem of obtaining and studying 
dimuonium, the bound state of the muon and anti-muon. Dimuonium has not yet 
been observed experimentally by anyone, and its first observation in 
experiments on the $\mu \mu$-tron would already be a significant discovery
\cite{2,3}. However, it seems that the first physical results on the 
$\mu \mu$-tron will be studies of the reactions $e^+e^-\to\mu^+\mu^-$ and 
$e^+e^-\to\mu^+\mu^-\gamma$ near the threshold. The former process can be used
for the precise measurement of the CMS energy and its spread at the 
$\mu \mu$-tron collider \cite{4}. Several accelerator technologies studies, 
like the development of a muon collider, could also benefit from these 
measurements \cite{5}.

There are several Monte Carlo programs that generate events of the process 
$e^+ e^- \to \mu^+ \mu^- \gamma$. Not public event generator AFKQED 
(supplemented by PHOTOS \cite{PHOTOS1,PHOTOS2} for the description of 
additional undetected final state radiation photons), based on the formulas 
of \cite{AFKM} in the case of $e^+ e^- \to \mu^+ \mu^- \gamma$, 
was used by BABAR collaboration in their analysis of hadron production via 
radiative return processes \cite{BABAR1,BABAR2}\footnote{The radiative return 
method is of great importance for measuring the hadron cross sections 
\cite{RR}. It is based on the observation that for radiation in the initial 
state, the following factorization is exact \cite{RR,RR1}: 
$$\frac{d\sigma(e^+e^-\to \mathrm{hadrons}+\gamma)(s,x)}{dx\,d\cos{\theta}}=
H(x,\cos{\theta})\;\sigma(e^+e^-\to \mathrm{hadrons})(s(1-x)),$$
where $x$ is the fraction of the beam energy taken by the radiative photon, 
and $\theta$ is the photon emission angle with respect to the beam. 
Therefore, since the function $H(x,\cos{\theta})$ is known from perturbative
QED, studying the reaction $e^+e^-\to \mathrm{hadrons}+\gamma$ will allow
to extract hadronic cross sections $\sigma(e^+e^-\to \mathrm{hadrons})$ in
a wide range of center-of-mass energies from the fixed nominal energy of the 
collider down to the $e^+e^-\to \mathrm{hadrons}$ production threshold. The 
only requirement is a high luminosity of the collider at nominal energy to 
compesate for the higher perturbative order of $e^+e^-\to \mathrm{hadrons}+
\gamma$ compared to $e^+e^-\to \mathrm{hadrons}$.}. PHOKHARA generator 
(beginning with version 4.0) \cite{PHOKHARA} includes next-to-leading order 
radiative corrections (one-loop corrections and emission of the second real 
photon) to the reaction $e^+ e^- \to \mu^+ \mu^- \gamma$. While PHOKHARA is 
based on fixed order perturbation theory calculations, KKMC generator 
\cite{KKMC1,KKMC2} uses Yennie-Frautschi-Suura type exponentiation formula 
to sum leading higher order effects in initial state radiation and thus
includes most of the third order leading-logarithmic contributions absent
in PHOKHARA 4.0. Let us mention also MCGPJ generator \cite{MCGPJ} in which
$e^+ e^- \to \mu^+ \mu^- \gamma$ is implemented using formulas of \cite{AFKM} 
and the formalism of structure functions is used to incorporate the leading 
logarithmic contributions related to the emission of photon jets in the 
collinear region.  

The aforementioned generators are more than enough for background studies 
at $\mu \mu$-tron. However, we decided to develop our own MUMUG generator 
mainly for pedagogical purposes -- to teach our students how to create a Monte
Carlo generator from scratch, and how to use the spinor helicity amplitudes 
method \cite{6}, which allows to organize the program in such a way that adding 
any new corrections (for example, Coulomb corrections important near the
threshold \cite{7}) becomes relatively easy.  This last feature is due to 
the fact that in the Monte Carlo generator based on the helicity amplitude 
method, we calculate induvidual helicity amplitudes as complex numbers in the 
corresponding software modules. Adding new contributions (comming, for example,
from the $Z$-boson exchange) will only require the addition of new program 
units to calculate these contributions, which are then added to the 
corresponding helicity amplitudes. At the same time, other program blocks that 
calculate the module square of the total amplitude and generate events based 
on it do not need to be changed.

At present MUMUG simulates $e^+ e^- \to \mu^+ \mu^- \gamma$ process only at the 
leading order (LO). To be able to perform internal consistency checks, various
methods is incorporated in MUMUG to obtain the tree level amplitudes needed
for the Monte Carlo simulation. In addition to the internal consistency checks,
MUMUG was tested against AFKQED in the BABAR energy region and good agreement 
has been reached. 

Much attention was paid to the optimization of the Monte Carlo algorithm, and
thanks to this MUMUG is a very fast LO generator (at BABAR energies it was 
more than two orders of magnitude faster than AFKQED).     

Below we give a very detailed description of the calculation of the 
tree-level amplitudes of $e^+ e^- \to \mu^+ \mu^- \gamma$  using the spinor 
technique, as well as of the Monte Carlo algorithm. We hope that this 
description will be of pedagogical value for students studying high energy 
physics.

The manuscript is organized as follows. In the next section, we outline the 
basics of the helicity amplitude method. After a brief historical 
introduction,  we present some of the relevant formulas for  the helicity 
amplitude method. After explaining how all spinors can be defined in terms of 
an auxiliary massless spinor with negative helicity, we present identities 
that are useful in calculating matrix elements. Then the second important 
ingredient of the helicity amplitude method is discussed, namely, the 
convenient choice of the four-vector of photon polarization. Some spinor 
products, the so-called $Z$-functions are introduced as useful building blocks 
for helicity amplitudes. The developed technique is used in the mext section
to calculate $e^+ e^- \to\mu^+ \mu^- \gamma$ tree level helicity amplitudes.
The fourth section describes the Monte Carlo algorithm  used in the event 
generator. After some general remarks about the acceptance-rejection
Monte Carlo method, the rough distributions for the initial and final state 
radiations are presented, which are used in the acceptance-rejection algorithm.
Brief concluding remarks end the article. For the convenience of readers, the 
appendix contains the Feynman rules used in the main text, as well as 
instructions for using the developed Fortran program. 

\section{Basics of the spinor technique}
In this section we shall present a summary of phase conventions and other 
relevant formulas of the spinor technique. More details can be found in the 
literature \cite{KS,ngamma,Xu,GPS} on which our presentation is based.  

\subsection{A brief historical introduction}
Experimental developments in high-energy physics around the 1980s showed 
that the theoretical description of higher order QCD reactions, in particular, 
hard processes with multi-jet production, requires improved computing 
technology, since several hundred Feynman diagrams could not be calculated 
by traditional methods, even with the use of symbolic-algebraic and numerical 
methods \cite{Mangano}. 

The usefulness of helicity amplitudes for the calculation of multi-parton 
scattering in the high-energy limit was recognized already in 1966-67
in pioneering papers by Bjorken and Chen, and by Reading-Henry \cite{Mangano}.
However, a real breakthrough in solving computational problems has  been 
provided by the development of the CALKUL collaboration in their classical
papers \cite{KS,ngamma,Xu,GPS} (see also other relevant references in 
\cite{Mangano,Dreiner}).

In the standard method for calculating cross sections of reactions with 
fermions in the initial and final states, we module square the corresponding
$S$-matrix amplitude described by several Feynman diagrams, sum and average 
(if necessary) over the spin states of the participating fermions, and then 
compute the traces of the products of the gamma matrices.
  
However, when the number of particles participating in the reaction is large,
the standard technique becomes impractical because the number of interfering 
Feynman diagrams increases rapidly.

One of the alternative methods that make calculations manageable in this case 
is the helicity amplitude method. In this method, the individual amplitudes 
that correspond to the scattering of the helicity eigenstates are calculated 
analytically as a complex number depending on the Lorentz scalars composed of 
the four-momenta of the participating particles. Since different helicity 
configurations do not interfere,  knowing the individual helicity amplitudes 
is sufficient to calculate the module square of the $S$-matrix amplitude by 
summing the modular squares of all contributing helicity amplitudes, and 
these calculations can be performed numerically on a computer. 

In fact, it is natural to use two-component Weyl-van der Waerden spinors in 
the helicity amplitude method, and a brief historical overview of the use of 
spinor techniques with the relevant literature can be found in \cite{Dreiner}.  

In this article we use the original variant of the helicity amplitude method,
as developed by the CALKUL collaboration \cite{6,KS,ngamma,Xu,GPS}. This 
technique turned out to be  extremely useful for QED calculations with 
massless fermions. However, its use in non-abelian theories such as  QCD 
required further refinements described in the review articles 
\cite{Peskin,Dixon}. Using the idea that on-shell one-loop amplitudes can be
reconstructed from their unitarity cuts, the use of the helicity amplitude
technique can also be extended to one-loop QCD calculations 
\cite{Melnikov,Huang}. However, these important developments are beyond the
scope of the present article.
      
In the next subsection we describe how both massless and massive spinors 
can be represented in terms of one auxiliary negative-helicity massless
spinor. In this way all spinor products are expressed in terms of two types 
of basic spinor inner products.

After providing some usefull spinor identities, we use gauge invariance of QED
to express the photon polarization four-vector in terms of a lightlike 
four-vector not collinear to the photon four-momentum or to the four-momentum 
of the auxiliary negative-helicity massless spinor mentioned previously. In 
specific calculations, a judicious choice of this auxiliary four-vector 
nulifies some Feynman diagrams and thus simplifies the calculation.

When massive spinors are involved, as in the case of $e^+ e^- \to\mu^+ \mu^- 
\gamma$, for a numerical evaluation it is convenient to express the helicity 
amplitudes in terms of  the so-called $Z$-functions, which we discuss in the 
last subsection.
 
\subsection{Massless and massive spinors}
All spinors are determined using an auxiliary negative-helicity massless
spinor $u_-(\xi)$ with a lightlike four-momentum $\xi$. The conditions 
determining $u_-(\xi)$ are as follows:
\begin{equation}
\hat \xi u_-(\xi)=0,\;\; \omega_- u_-(\xi)=u_-(\xi),\;\;
\omega_\lambda=\frac{1}{2}(1+\lambda \gamma_5),\;\; \lambda=\pm 1,
\label{nhbs} \end{equation}
and its normalization condition is $u_-(\xi) {\bar u}_ -(\xi)=
\omega_- \hat \xi$ (we use conventions for the space-time metric etc. 
accepted in the modern QFT textbooks \cite{QFT}. In particular, $\hat \xi=
\xi_\mu\gamma^\mu$).

Under space inversion $u_-(\xi) \to \gamma_0 u_-(\xi)\sim u_+(\xi^\prime)$,
where $\xi^\prime=(\xi_0,-\vec{\xi}~)$ and $\sim$ means equality up to a phase. 
Let $\eta=(0,\vec{n}~),\;\vec{n}^2=1$ and $\vec{n}\cdot \vec{\xi}=0$. 
Then $\hat R_{\vec{n}}(\pi)\xi^\prime=\xi$, where $\hat R_{\vec{n}}(\phi)$ stands 
for a rotation by the angle $\phi$ around the axis $\vec{n}$. Therefore 
$\hat R_{\vec{n}}(\pi)\gamma_0 u_-(\xi)\sim u_+(\xi)$. But 
$$\hat R_{\vec{n}}(\pi)\gamma_0 u_-(\xi)=\exp{\left (-i\frac{\pi}{2}\,\vec{n}
\cdot\vec{\Sigma}\right )}\gamma_0 u_-(\xi) = -i\vec{n}\cdot\vec{\Sigma}\,
\gamma_0 u_-(\xi),$$
and $-i\vec{n}\cdot\vec{\Sigma}\,\gamma_0u_-(\xi)=
i\vec{n}\cdot\vec{\Sigma}\,\gamma_0\gamma_5 u_-(\xi)$. For  gamma matrices
we use chiral representation  
$$
\gamma_0=\left (\begin{array}{cc} 0 & 1 \\ 1 & 0 \end{array}\right ), \; \;
\gamma^k=\left (\begin{array}{cc} 0 & -\sigma_k \\ \sigma_k & 0 \end{array}
\right ), \;\; 
\gamma_5=\left (\begin{array}{cc} 1 & 0 \\ 0 & -1 \end{array}
\right ), \;\; \Sigma^k=\left (\begin{array}{cc} \sigma_k & 0 \\ 0 & \sigma_k
\end{array}\right ),
$$
so it can be checked easily that $\Sigma^k\gamma_0\gamma_5=\gamma^k$. 
Therefore $u_+(\xi)\sim i n^k \gamma^k
u_-(\xi)=-i\hat \eta u_-(\xi)$. As was already mentioned, this relation 
determines the positive-helicity massless spinor $u_+(\xi)$ only up to 
a phase, and the phase is fixed by the following Kleiss-Stirling convention 
\cite{KS}
\begin{equation}
u_+(\xi)=\hat \eta u_-(\xi),\;\; \eta^2=-1, \;\; \eta \cdot \xi =0.
\label{phbs} \end{equation}
It should be mentioned that the modern literature \cite{Peskin,Dixon} uses 
a slightly different way of introducing the basic massless spinors and the 
associated phase convention. 

Now let us construct a general massive spinor with four-momentum $p$ ($p^2=
m^2$) which besides the Dirac equation $(\hat p -m) u_\lambda (p)=0$ satisfies
the normalization condition
$$u_\lambda (p){\bar u}_\lambda (p)=\frac{1}{2}(1+\lambda \gamma_5 \hat s)
(\hat p +m),$$
where $s$ is the spin quantization vector with properties $s^2=-1,
\;\; s\cdot p=0$. This can be done by noting that
$$(\hat p +m)u_\lambda (\xi)= \left [u_\lambda (p){\bar u}_\lambda (p)+
 u_{-\lambda} (p){\bar u}_{-\lambda} (p) \right ] u_\lambda (\xi)= $$ 
\begin{equation}
\left [{\bar u}_\lambda (p)u_\lambda (\xi)\right ] u_\lambda (p)+
\left [{\bar u}_{- \lambda} (p)u_\lambda (\xi)\right ] u_{- \lambda} (p).
\label{spindef} \end{equation}
Now
\begin{equation}
\left |{\bar u}_\lambda (p)u_\lambda (\xi)\right |^2=Sp\left \{\frac{1}{2}
(1+\lambda\gamma_5\hat s)(\hat p +m)\omega_\lambda \hat \xi\right \}=
p\cdot \xi-ms\cdot\xi.\label{coefl} \end{equation}
Analogously
\begin{equation}
\left |{\bar u}_{-\lambda} (p)u_\lambda (\xi)\right |^2=p\cdot \xi+
ms\cdot\xi. \label{coefml} \end{equation}
But $(p-ms)^2=0$, so we may try $p-ms=k\xi$ and then $p\cdot s=0$ determines
the coefficient $k=m^2/p\cdot\xi$. Therefore the following choice is a valid
spin quantization vector:
$$s=\frac{p}{m}-\frac{m}{p\cdot \xi}\,\xi.$$
With this spin quantization vector $ms\cdot\xi=p\cdot\xi$ and we get from 
(\ref{spindef}) (the phase is again fixed according to Kleiss-Stirling 
\cite{KS})
\begin{equation}
u_\lambda (p)=\frac{(\hat p +m)u_{-\lambda} (\xi)}{\sqrt{2p\cdot \xi}}.
\label{uspin} \end{equation}
For the antiparticle spinor we have $(\hat p +m)v_\lambda (p)=0, \;\;
v_\lambda (p){\hat v}_\lambda (p)=\frac{1}{2}(1+\lambda \gamma_5 \hat s)
(\hat p -m)$. So it can be derived in the same way as the u-type spinor above 
but by the change $m\to -m$. Note that (\ref{spindef}), (\ref{coefl}) and 
(\ref{coefml}) indicate that one should also change $\lambda \to -\lambda$. 
Therefore a general antiparticle spinor is
\begin{equation}
v_\lambda (p)=\frac{(\hat p -m)u_\lambda (\xi)}{\sqrt{2p\cdot \xi}}.
\label{vspin} \end{equation}

\noindent Note that
$$\frac{\hat p u_{-\lambda}(\xi)}{\sqrt{2p\cdot\xi}}= \frac{\hat p_\xi 
u_{-\lambda}(\xi)}{\sqrt{2p_\xi\cdot\xi}}=u_\lambda (p_\xi),$$
where $p_\xi$ is the lightlike four-vector:
$$p_\xi=p-\frac{m^2}{2p\cdot\xi}\xi,\;\; p_\xi^2=0.$$
Therefore the massive spinors can always be decomposed in terms of massless 
spinors as follows
\begin{equation}
u_\lambda (p)=u_\lambda (p_\xi)+\frac{m}{\sqrt{2p\cdot\xi}}u_{-\lambda} 
(\xi), \;\; v_\lambda (p)=u_{-\lambda} (p_\xi)-\frac{m}{\sqrt{2p\cdot\xi}}
u_\lambda (\xi).
\label{decomp} \end{equation}

\subsection{Useful spinor identities}
In QCD calculations, the amplitudes have a large number of momenta, and for 
massless spinors it is useful to use the shorthand notation 
\cite{Peskin,Dixon,Xu} (not to be confused with Dirac's bra-ket notation for
quantum state vectors)
\begin{equation}
\begin{aligned}
&u_-(\xi)=|\xi]\equiv |\xi -\rangle,\;\;u_+(\xi)=|\xi\rangle\equiv 
|\xi +\rangle, \\
&\bar{u}_-(\xi)=\langle\xi|\equiv \langle\xi-|,\;\;
\bar{u}_+(\xi)=[\xi|\equiv\langle \xi+|.
\end{aligned}
\label{brakets}
\end{equation}
Although we don't really need these notations for $e^+e^-\to\mu^+\mu^-\gamma$, 
they are now standard. Therefore, it is important for students to get to know 
them by reading modern review literature \cite{Peskin,Dixon,Melnikov}.
 
For massless spinors the following identities proved to be useful \cite{KS}.
The reversion identity:
\begin{equation}
{\bar u}_{\lambda_1}(p_1)\Gamma u_{\lambda_2}(p_2)=\lambda_1\lambda_2
{\bar u}_{-\lambda_2}(p_2)\Gamma^R u_{-\lambda_1}(p_1),
\label{reverseid} \end{equation}
where $\Gamma$ stands for any string of Dirac matrices and $\Gamma^R$ is the
same string written in reversed order.

\noindent The Chisholm identity:
\begin{equation}
\left \{ {\bar u}_\lambda(p_1)\gamma^\mu u_\lambda(p_2)\right \}\gamma_\mu=
2u_\lambda(p_2){\bar u}_\lambda(p_1)+2u_{-\lambda}(p_1){\bar u}_{-\lambda}
(p_2). \label{Chisholmid} \end{equation}

These identities allow to get expression for any amplitude in terms of 
spinor inner products. For massless spinors the basic inner products are
\begin{equation}
\begin{aligned}
&[p_1p_2]=s(p_1,p_2)={\bar u}_+(p_1)u_-(p_2)=-s(p_2,p_1),\\
&\langle p_1p_2\rangle=t(p_1,p_2)={\bar u}_-(p_1)u_+(p_2)=[s(p_2,p_1)]^*.
\end{aligned}
\label{binp} \end{equation}
It is not difficult to get explicit expression for $X=\sqrt{4p_1\cdot\xi 
~p_2\cdot\xi}~s(p_1,p_2)$:
$$X={\bar u}_-(\xi)\hat p_1\hat p_2
u_+(\xi)={\bar u}_-(\xi)\hat p_1\hat p_2\hat \eta u_-(\xi)=
Sp\left \{ \omega_-\hat\xi \hat p_1\hat p_2\hat \eta\right \}.$$
Therefore
$$s(p_1,p_2)=\frac{p_1\cdot\xi~p_2\cdot\eta-p_1\cdot\eta~p_2\cdot\xi-
i\epsilon_{\mu\nu\sigma\tau}\xi^\mu\eta^\nu p_1^\sigma p_2^\tau}
{\sqrt{p_1\cdot \xi~p_2\cdot \xi}}, \;\; \epsilon_{0123}=1.$$ 
Below we shall make Kleiss-Stirling \cite{KS} choice for the auxiliary vectors
$\xi$ and $\eta$:
\begin{equation}
\xi=(1,1,0,0),\;\;\; \eta=(0,0,1,0).
\label{xieta} \end{equation}
Then
\begin{equation}
s(p_1,p_2)=\left ( p_{1y}+ip_{1z}\right )\sqrt{\frac{p_{20}-p_{2x}}
{p_{10}-p_{1x}}}-\left ( p_{2y}+ip_{2z}\right )\sqrt{\frac{p_{10}-p_{1x}}
{p_{20}-p_{2x}}}. \label{sp1p2} \end{equation}
This formula is valid for any lightlike momenta $p_1,p_2$ not collinear
to $\xi$. The same is true for other formulas given above also. The case when 
one of the four-momenta is collinear to $\xi$ have to be treated separately. 
In particular, we have 
$$ s(p,\xi)=s(\xi,p)=\sqrt{2 p\cdot \xi}, $$
for any lightlike four-momentum $p$.

Decomposition (\ref{decomp}) allows to evaluate inner products
$$s_{\lambda_1,\lambda_2}(p_1,p_2)={\bar u}_{\lambda_1}
(p_1)u_{\lambda_2}(p_2)$$
for massive spinors. In particular $s_{+-}(p_1,p_2)=-s_{+-}(p_2,p_1)$
is given by the very formula (\ref{sp1p2}) and
$s_{-+}(p_1,p_2)=-[s_{+-}(p_1,p_2)]^*$. While 
the equal helicity inner products depend explicitly on the masses (of course,
for massive particles helicity is not a Lorentz-invariant concept. 
nevertheless we shall use this terminology in massive case too as some useful 
convention):
$$
s_{++}(p_1,p_2)=s_{++}(p_2,p_1)=s_{--}(p_1,p_2)=$$
\begin{equation}
m_1\sqrt{\frac{p_2\cdot\xi}
{p_1\cdot\xi}}+m_2\sqrt{\frac{p_1\cdot\xi}{p_2\cdot\xi}}=
m_1\sqrt{\frac{p_{20}-p_{2x}}{p_{10}-p_{1x}}}+
m_2\sqrt{\frac{p_{10}-p_{1x}}{p_{20}-p_{2x}}}.
\label{sppmm} \end{equation}
Inner products for antiparticle spinors are obtained by the following simple
rule (as (\ref{decomp}) indicates): substitute $m\to -m,\;\; \lambda \to
-\lambda$ into the rhs of (\ref{sp1p2}) and (\ref{sppmm}) when in the lhs 
$u_\lambda(p),\; p^2=m^2$ spinor is replaced by $v_\lambda(p)$ antiparticle
spinor.

It turns out \cite{KM} that the most numerically stable way of computing
the scalar product $p_1\cdot p_2$, especially when the masses are zero or
small, is to use the formula
\begin{equation}
p_1\cdot p_2=\frac{1}{4}\sum\limits_{\lambda_1,\lambda_2} \left |
s_{\lambda_1,\lambda_2}(p_1,p_2)\right |^2-m_1m_2,
\label{p1p2} \end{equation} 
which can be proved by using the explicit forms for $s_{\lambda_1,\lambda_2}
(p_1,p_2)$ as given by (\ref{sp1p2}) and (\ref{sppmm}).

To end our discussion of spinors, let us mention another interesting identity
(the Schouten identity):
\begin{equation}
s_{+-}(p_1,p_2)s_{+-}(p_3,p_4)+s_{+-}(p_2,p_3)s_{+-}(p_1,p_4)+
s_{+-}(p_3,p_1)s_{+-}(p_2,p_4)=0.
\label{Schoutenid} \end{equation}
The most easy way to check it is to use (\ref{sp1p2}) and some computer 
program for algebraic manipulations, for example REDUCE \cite{REDUCE}.

\subsection{Photon polarization four-vectors}
The next important ingredient of the spinor technique under discussion is
a convenient choice of the photon polarization four-vector. This polarization 
four-vector $\epsilon_\lambda^\mu$ corresponding to a state of definite 
helicity $\lambda=\pm 1$ should satisfy the equations 
\begin{equation}
\epsilon_\lambda\cdot k=0,\;\; \epsilon_\lambda\cdot\epsilon_\lambda=0,\;\;
\epsilon^*_\lambda\cdot\epsilon_\lambda=-1,
\label{epseq} \end{equation}   
$k$ being the photon momentum. The following choice proved to be useful:
\begin{equation}
\left (\epsilon_\lambda^\mu(k;p)\right )^*=\frac{{\bar u}_\lambda (k)
\gamma^\mu u_\lambda (p)}{\sqrt{2} s_{-\lambda,\lambda}(k,p)}.
\label{epsilon} \end{equation}
Here $p$ is some auxiliary lightlike four-vector not collinear to $k$ or
$\xi$. It can be checked that (\ref{epsilon}) indeed satisfies relations
(\ref{epseq}). For example, using
$$\left |s_{-\lambda,\lambda}(k,p)\right |^2=2k\cdot p$$
and
$${\bar u}_\lambda (k)\gamma^\mu u_\lambda (p) {\bar u}_\lambda (p)
\gamma_\mu u_\lambda (k)=Sp\left \{ \omega_\lambda \hat k \gamma^\mu
\omega_\lambda\hat p \gamma_\mu\right \}=
Sp\left \{ \omega_\lambda \hat k \gamma^\mu\hat p \gamma_\mu\right \}=  
-4k\cdot p$$
we get $\epsilon^*_\lambda(k;p)\cdot \epsilon_\lambda(k;p)=-1$.
Following \cite{KKMC2}, we keep in (\ref{epsilon}) explicit complex conjugation
which for outgoing photons is canceled by another conjugation required by
the Feynman rules.

Any other four-vector satisfying (\ref{epseq}) is as good as (\ref{epsilon})
itself to represent the photon polarization and therefore can differ from 
(\ref{epsilon}) only by a phase and some gauge transformation. In particular,
for another choice of the auxiliary four-vector we will have
$$\epsilon_\lambda^\mu(k;q)=e^{i\Phi(q,p)}\epsilon_\lambda^\mu(k;p)+
\beta_\lambda (q,p,k)k_\mu.$$
Let us calculate the phase $\Phi(q,p)$.
$$e^{i\Phi(q,p)}=-\epsilon^*_\lambda (k,p)\cdot \epsilon_\lambda (k,q)=
-\frac{{\bar u}_\lambda (k)\gamma^\mu u_\lambda (p){\bar u}_\lambda (q)
\gamma_\mu u_\lambda (k)}{2s_{-\lambda,\lambda}(k,p)s^*_{-\lambda,\lambda}
(k,q)}.$$
Using the Chisholm identity (\ref{Chisholmid}) the numerator can be rewritten
as follows
$${\bar u}_\lambda (k)\left \{{\bar u}_\lambda (q)\gamma_\mu u_\lambda (k)
\right \}\gamma^\mu u_\lambda (p)=$$ 
$$2{\bar u}_\lambda (k)\left [u_{-\lambda}
(q){\bar u}_{-\lambda}(k)+u_\lambda (k) {\bar u}_\lambda (q) \right ]
u_\lambda (p)=2s_{\lambda,-\lambda}(k,q)s_{-\lambda,\lambda}(k,p).$$
But $s_{\lambda,-\lambda}(k,q)=-s^*_{-\lambda,\lambda}(k,q)$ and we get
$$e^{i\Phi(q,p)}=1.$$
So 
\begin{equation}
\epsilon_\lambda^\mu(k;q)=\epsilon_\lambda^\mu(k;p)+
\beta_\lambda(q,p,k)k_\mu
\label{epsqp} \end{equation} 
and for each gauge invariant subset of diagrams we can
choice the auxiliary four-vector $p$ completely arbitrarily without generating
relative complex phases -- certainly a nice property. 

Using the Chisholm identity again we get
\begin{equation}
\hat \epsilon^*_\lambda (k;p)=\frac{\sqrt{2}}{s_{-\lambda,\lambda}(k,p)}
\left [u_\lambda(p){\bar u}_\lambda (k)+u_{-\lambda}(k){\bar u}_
{-\lambda} (p)\right ].
\label{hateps} \end{equation}

\noindent From this relation the following ``magic'' identities follow 
\begin{equation}
\hat \epsilon ^*_\lambda (k;p)u_\lambda (p)=\hat \epsilon ^*_\lambda (k;p)
v_{-\lambda} (p)={\bar v}_\lambda (p)\hat \epsilon^*_\lambda (k;p)=
{\bar u}_{-\lambda} (p)\hat \epsilon^*_\lambda (k;p)=0.
\label{magicid} \end{equation}
\noindent
In fact our polarization four-vector (\ref{epsilon}) corresponds to the axial 
gauge $p\cdot\epsilon_\lambda=0$ and for the photon polarization sum we have
\begin{equation}
\sum\limits_\lambda \epsilon_\lambda^\mu (k;p)\epsilon_\lambda^{*\nu} (k;p)=
-g^{\mu\nu}+\frac{p^\mu k^\nu+p^\nu k^\mu}{p\cdot k}.
\label{phpolsum} \end{equation}
Indeed
$$\sum\limits_\lambda \epsilon_\lambda^\mu \epsilon_\lambda^{*\nu} =
\frac{\sum\limits_\lambda {\bar u}_\lambda (p)\gamma^\mu u_\lambda (k) 
{\bar u}_\lambda (k) \gamma^\nu u_\lambda (p)}{4k\cdot p}=
\frac{\sum\limits_\lambda Sp \left \{\omega_\lambda \hat p \gamma^\mu
\omega_\lambda \hat k \gamma^\nu \right \}}{4k\cdot p}=$$ $$
\frac{Sp \left \{\sum\limits_\lambda \omega_\lambda \hat p \gamma^\mu
\hat k \gamma^\nu \right \}}{4k\cdot p}=\frac{Sp \left \{\hat p \gamma^\mu 
\hat k \gamma^\nu \right \}}{4k\cdot p}=-g^{\mu\nu}+\frac{k^\mu p^\nu+k^\nu 
p^\mu}{p\cdot k}.$$
\noindent
Substituting $\hat p=u_\lambda (p) {\bar u}_\lambda (p)+
u_{-\lambda }(p) {\bar u}_{-\lambda} (p)$ in
$$p\cdot \epsilon^*_\lambda (k;q)=\frac{{\bar u}_\lambda (k)\hat p u_\lambda
(q)}{\sqrt{2}s_{-\lambda,\lambda}(k,q)}$$
we get
\begin{equation}
p\cdot \epsilon^*_\lambda (k;q)=\frac{s_{\lambda,-\lambda}(k,p)
s_{-\lambda,\lambda}(p,q)}{\sqrt{2}s_{-\lambda,\lambda}(k,q)}.
\label{pepsq}\end{equation}
In particular
\begin{equation}
p\cdot \epsilon^*_\lambda (k;p)=k\cdot \epsilon^*_\lambda (k;p)=0.
\label{pepsp} \end{equation}
\noindent
The relation (\ref{pepsq}) allows to calculate $\beta_\lambda(q,p,k)$
in (\ref{epsqp}):
\begin{equation}
\beta_\lambda(q,p,k)=\frac{\sqrt{2}s_{-\lambda,\lambda}(p,q)}
{s_{-\lambda,\lambda}(p,k)s_{-\lambda,\lambda}(k,q)}.
\label{beta} \end{equation}

\subsection{$Z$-functions as building blocks for helicity amplitudes}
The following $Z$-functions are useful \cite{MMM,BMM} building blocks for 
helicity amplitudes:
\begin{equation}
Z_{\lambda_1 ,\lambda_2,\lambda_3,\lambda_4}^{\,\epsilon_1,\,\epsilon_2,\,
\epsilon_3, \,
\epsilon_4}(p_1,p_2,p_3,p_4)={\bar u}_{\lambda_1}(p_1)\gamma_\mu
u_{\lambda_2}(p_2){\bar u}_{\lambda_3}(p_3)\gamma^\mu
u_{\lambda_4}(p_4),
\label{Zdef} \end{equation}
where it is assumed that every mass in this expression is written in the form
$\epsilon_i m_i$, with $\epsilon_i=0,\pm 1$, to ensure a simple use of the  
$\lambda\to -\lambda,\;\; m\to -m$ substitution rule for antiparticle spinors. 
Decomposition (\ref{decomp}) of massive spinors in terms of massless ones 
allows to express $Z$-functions trough spinor inner products. The calculation
is straightforward although rather lengthy. To present the results let us
introduce for a moment a shorthand notations 
$$Z(\lambda_1,\lambda_2,\lambda_3,\lambda_4)\equiv 
Z_{\lambda_1 ,\lambda_2,\lambda_3,\lambda_4}^{\,\epsilon_1,\,\epsilon_2,\,
\epsilon_3, \, \epsilon_4}(p_1,p_2,p_3,p_4),\;\;\chi_i=\sqrt{2p_i\cdot\xi},
\;\;, \mu_i=\frac{\epsilon_i m_i}{\chi_i}.$$
Then for half of the possible helicity configurations we have
\begin{eqnarray} 
Z(+,+,+,+)&=&-2\left [s_{+-}(p_1,p_3)s_{-+}(p_2,p_4)-\mu_1\mu_2
\chi_3\chi_4-\mu_3\mu_4\chi_1\chi_2\right ],  \nonumber \\  
Z(+,+,+,-)&=&2\chi_2\left [\mu_3s_{+-}(p_1,p_4)-
\mu_4s_{+-}(p_1,p_3) \right ],  \nonumber \\ 
Z(+,+,-,+)&=&2\chi_1\left [\mu_3s_{-+}(p_2,p_4)-
\mu_4s_{-+}(p_2,p_3) \right ],  \nonumber \\ 
Z(+,+,-,-)&=&-2\left [s_{+-}(p_1,p_4)s_{-+}(p_2,p_3)-
\mu_1\mu_2\chi_3\chi_4-\mu_3\mu_4\chi_1\chi_2\right ],  \nonumber \\
Z(+,-,+,+)&=&2\chi_4\left [\mu_2s_{+-}(p_1,p_3)-
\mu_1s_{+-}(p_2,p_3) \right ], \label{Zexpr} \\
Z(+,-,+,-)&=& 0, \nonumber \\
Z(+,-,-,+)&=& 2\left [\mu_1\mu_3\chi_2\chi_4+\mu_2\mu_4\chi_1\chi_3-
\mu_1\mu_4\chi_2\chi_3-\mu_2\mu_3\chi_1\chi_4\right ], \nonumber \\
Z(+,-,-,-)&=& 2\chi_3\left [\mu_2s_{+-}(p_1,p_4)-
\mu_1s_{+-}(p_2,p_4) \right ]. \nonumber 
\end{eqnarray} 
The remaining half can be obtained by exchanging $+ \leftrightarrow -$
in the above expressions. 

For illustration purposes let us sketch the derivation of $Z(+,+,+,+)$.
Using the decomposition (\ref{decomp}) we get (note that 
${\bar u}_\pm(p)\gamma_\mu u_\mp(q)=0$ for lightlike
four-momenta $p$ and $q$ because
for such momenta $u_\pm$ are chirality eigenstates)
$$Z(+,+,+,+)=\left [ {\bar u}_+(p_{1\xi})\gamma_\mu u_+(p_{2\xi})+
\mu_1\mu_2{\bar u}_-(\xi)\gamma_\mu u_-(\xi)\right]\times $$
$$ \left[
{\bar u}_+(p_{3\xi})\gamma^\mu u_+(p_{4\xi})+
\mu_3\mu_4{\bar u}_-(\xi)\gamma^\mu u_-(\xi)\right].$$
For lightlike four-momenta (but not for $\xi$ which plays a special role
in our construction \cite{KKMC2}) we can enjoy the Chisholm identity 
(\ref{Chisholmid}) and obtain
$${\bar u}_+(p_{1\xi})\gamma_\mu u_+(p_{2\xi})
{\bar u}_+(p_{3\xi})\gamma^\mu u_+(p_{4\xi})= $$ $$
2{\bar u}_+(p_{3\xi})\left [u_+(p_{2\xi}){\bar u}_+(p_{1\xi})+
u_-(p_{1\xi}){\bar u}_-(p_{2\xi})\right ]u_+(p_{4\xi})=$$
$$2s_{+-}(p_{3\xi},p_{1\xi})s_{-+}(p_{2\xi},p_{4\xi})=
-2s_{+-}(p_1,p_3)s_{-+}(p_2,p_4).$$
The last step follows from $s_{\pm\mp}(p_\xi,q_\xi)=s_{\pm\mp}(p,q)=-
s_{\pm\mp}(q,p)$. Analogously
$${\bar u}_+(p_{1\xi})\gamma_\mu u_+(p_{2\xi})
{\bar u}_-(\xi)\gamma^\mu u_-(\xi)=$$ $$
2{\bar u}_-(\xi)\left [u_+(p_{2\xi}){\bar u}_+(p_{1\xi})+
u_-(p_{1\xi}){\bar u}_-(p_{2\xi})\right ]u_-(\xi)=$$
$$2s_{-+}(\xi,p_{2\xi})s_{+-}(p_{1\xi},\xi)=2\chi_2\chi_1.$$
At last, because $\xi^2=0$, we have
$${\bar u}_-(\xi)\gamma_\mu u_-(\xi)
{\bar u}_-(\xi)\gamma^\mu u_-(\xi)=$$ $$
Sp\left \{\omega_-\hat \xi\gamma_\mu\omega_-\hat \xi\gamma^\mu\right \}=
Sp\left \{\omega_-\hat \xi\gamma_\mu\hat \xi\gamma^\mu\right \}=
-2Sp\left \{ \omega_-\hat \xi\hat \xi \right \}=0.$$
Putting all pieces together, hence the first line of (\ref{Zexpr}) follows. 
 
\section{Tree level helicity amplitudes}
Feynman diagrams for the process $e^-(p_-)+e^+(p_+)\to \mu^-(q_-)+\mu^+(q_+)
+\gamma (k)$ at tree level are naturally divided into two gauge-invariant
subsets: initial state radiation and final state radiation. 

\subsection{Initial state radiation}
Let us consider the initial state radiation first. The helicity amplitude 
method gives especially simple and elegant results in massless case. For our 
radiative return studies we can neglect the electron mass assuming that the 
tagged hard photon is not emitted at very small angles. But unfortunately we 
can not neglect the muon mass if we are interested in production of the 
muon-antimuon pairs with sufficiently small invariant masses. For massless
fermions helicity is not changed by photon emission. That is the helicity
is conserved along initial electron line. Therefore the helicity amplitude
${\cal A}_{\lambda^e_-,-\lambda^e_+,\lambda^\mu_-,-\lambda^\mu_+,\lambda^\gamma}$ is zero if 
$\lambda^e_-=\lambda^e_+$ (note that the incoming positron with helicity 
$\lambda^e_+$ is equivalent to the outgoing electron with helicity 
$-\lambda^e_+$). So in our approximation of massless electron 
we have to consider only helicity amplitudes with opposite electron and 
positron helicities.

Let us consider in some details the calculation of ${\cal A}^{(ISR)}_{+,+,
\lambda_3,\lambda_4,+}$. Because of the ``magic'' identities (\ref{magicid}),
in this case it is convenient to choose the auxiliary vector $p$ in the 
photon polarization four-vector definition (\ref{epsilon}) to be the electron
momentum $p_-$. Then only one initial state radiation diagram will contribute.
Namely the one with the photon radiation from the positron:

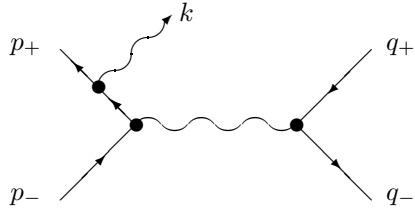
\begin{center}
\begin{picture}(20000,12000)
\drawline\fermion[\SE\REG](2000,9000)[2000]
\drawarrow[\NW\ATBASE](\pmidx,\pmidy)
\global\advance\pfrontx by -1900
\put(\pfrontx,\pfronty){$p_+$}
\put(\pbackx,\pbacky){\circle*{500}}
\drawline\photon[\NE\REG](\pbackx,\pbacky)[4]
\drawarrow[\NE\ATBASE](\pbackx,\pbacky)
\global\advance\pbackx by 500
\put(\pbackx,\pbacky){$k$}
\drawline\fermion[\SE\REG](\pfrontx,\pfronty)[2000]
\drawarrow[\NW\ATBASE](\pmidx,\pmidy)
\drawarrow[\NW\ATBASE](\pmidx,\pmidy)
\put(\pbackx,\pbacky){\circle*{500}}
\drawline\fermion[\SW\REG](\pbackx,\pbacky)[4000]
\global\advance\pbackx by -1900
\put(\pbackx,\pbacky){$p_-$}
\drawarrow[\NE\ATBASE](\pmidx,\pmidy)
\drawline\photon[\E\REG](\pfrontx,\pfronty)[6]
\drawline\fermion[\NE\REG](\photonbackx,\photonbacky)[4000]
\drawarrow[\SW\ATBASE](\pmidx,\pmidy)
\global\advance\pbackx by 500
\put(\pbackx,\pbacky){$q_+$}
\put(\pfrontx,\pfronty){\circle*{500}}
\drawline\fermion[\SE\REG](\photonbackx,\photonbacky)[4000]
\drawarrow[\SE\ATBASE](\pmidx,\pmidy)
\global\advance\pbackx by 500
\put(\pbackx,\pbacky){$q_-$}
\end{picture}
\end{center}

\noindent According to the Feynman rules it is straightforward to get
$${\cal A}=\frac{ie^3}{s^\prime}~\frac{1}{-2 p_+\cdot k}
{\bar v}_-(p_+)\hat \epsilon_+^*(k;p_-) (-\hat p_++\hat k)\gamma_\mu u_+(p_-)
{\bar u}_{\lambda_3}(q_-) \gamma^\mu v_{-\lambda_4} (q_+), $$
where $s^\prime=(q_-+q_+)^2$. Because ${\bar v}_-(p_+)\hat \epsilon_+^*(k;p_-)
\hat p_+=2p_+\cdot\epsilon_+^*(k;p_-){\bar v}_-(p_+)$, 
we will have ${\cal A}={\cal A}_1+{\cal A}_2$, where
$${\cal A}_1=\frac{ie^3}{s^\prime}~\frac{p_+\cdot\epsilon_+^*(k;p_-)}
{p_+\cdot k}{\bar v}_-(p_+)\gamma_\mu u_+(p_-){\bar u}_{\lambda_3}(q_-) 
\gamma^\mu v_{-\lambda_4} (q_+).$$
Using (\ref{pepsq}) and $2p_+\cdot k=-s_{+-}(k,p_+)s_{-+}(k,p_+)$, we get
$${\cal A}_1=\frac{i\sqrt{2}e^3}{s^\prime}~\frac{s_{-+}(p_-,p_+)}
{s{-+}(k,p_-) s{-+}(k,p_+)}Z_{+,+,\lambda_3,\,\lambda_4}
^{0\, ,\,0\, ,\,1\; ,-1}(p_+,p_-,q_-,q_+).$$
For ${\cal A}_2$ we have
$${\cal A}_2=\frac{ie^3}{s^\prime}~\frac{1}{-2p_+\cdot k} {\bar v}_-(p_+)
\hat \epsilon^*_+(k;p_-)\hat k \gamma_\mu u_+(p_-){\bar u}_{\lambda_3}(q_-)
\gamma^\mu v_{-\lambda_4} (q_+).$$
Now $\hat \epsilon^*_+(k;p_-)\hat k \gamma_\mu u_+(p_-)=-\hat k
\hat \epsilon^*_+(k;p_-) \gamma_\mu u_+(p_-)=-2\epsilon^*_{+\mu}(k;p_-)\hat k
u_+(p_-)$, because $k\cdot\epsilon^*_{+}(k;p_-)=0$ and $\hat \epsilon^*_+
(k;p_-)u_+(p_-)=0$. Therefore
$${\cal A}_2=\frac{ie^3}{s^\prime}~\frac{1}{p_+\cdot k}
{\bar v}_-(p_+)\hat k u_+(p_-){\bar u}_{\lambda_3}(q_-)\hat \epsilon^*_+
(k;p_-) v_{-\lambda_4} (q_+).$$
The first spinor factor is calculated by substituting ${\bar v}_-(p_+)=
{\bar u}_+(p_+)$ and $\hat k=u_+(k){\bar u}_+(k)+u_-(k){\bar u}_-(k)$. The 
result is 
$${\bar v}_-(p_+)\hat k u_+(p_-)=s_{+-}(p_+,k)s_{-+}(k,p_-).$$
For the second spinor factor we have
$${\bar u}_{\lambda_3}(q_-)\hat \epsilon^*_+(k;p_-) v_{-\lambda_4} (q_+)=$$
$$ \frac{{\bar u}_+(k)\gamma_\mu u_+(p_-){\bar u}_{\lambda_3}(q_-)\gamma^\mu
v_{-\lambda_4} (q_+)}{\sqrt{2}s_{-+}(k,p_-)}=
\frac{Z_{+,+,\lambda_3,\,\lambda_4}^{0\, ,\,0\,,\,1\; ,-1}
(k,p_-,q_-,q_+)}{\sqrt{2}s_{-+}(k,p_-)}.$$
Combining all pieces together, we get
$${\cal A}_2=-\frac{i\sqrt{2}e^3}{s^\prime}~\frac{s_{-+}(p_-,k)}
{s_{-+}(k,p_-) s_{-+}(k,p_+)}Z_{+,+,\lambda_3,\, \lambda_4}
^{0\, ,\,0\, ,\,1\; ,-1}(k,p_-,q_-,q_+)$$
and
$${\cal A}^{(ISR)}_{+,+,\lambda_3,\lambda_4,+}=\frac{i\sqrt{2}e^3}{s^\prime}
\times$$
\begin{equation}
\frac{s_{-+}(p_-,p_+)Z_{+,+,\lambda_3,\,\lambda_4}^{0\, ,\,0\, ,\, 1\; ,-1}
(p_+,p_-,q_-,q_+)-
s_{-+}(p_-,k)Z_{+,+,\lambda_3,\,\lambda_4}
^{0\, ,\,0\, ,\,1\; ,-1}(k,p_-,q_-,q_+)}
{s_{-+}(k,p_-) s_{-+}(k,p_+)}.
\label{AISRpmp} \end{equation}

For ${\cal A}^{(ISR)}_{+,+,\lambda_3,\lambda_4,-}$ it is useful to take 
$p=p_+$ as the auxiliary four-vector for the photon polarization. Then only
the following diagram will contribute:
   
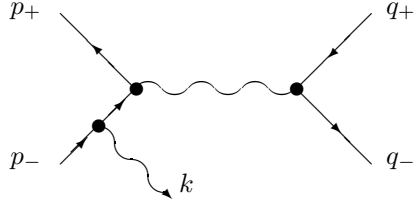
\begin{center}
\begin{picture}(20000,12000)
\drawline\fermion[\SE\REG](2000,9000)[4000]
\global\Xone=\pbackx
\global\Yone=\pbacky
\drawarrow[\NW\ATBASE](\pmidx,\pmidy)
\put(\pbackx,\pbacky){\circle*{500}}
\global\advance\pfrontx by -1900
\put(\pfrontx,\pfronty){$p_+$}
\drawline\fermion[\SW\REG](\pbackx,\pbacky)[2000]
\drawarrow[\NE\ATBASE](\pmidx,\pmidy)
\put(\pbackx,\pbacky){\circle*{500}}
\drawline\photon[\SE\REG](\pbackx,\pbacky)[4]
\drawarrow[\SE\ATBASE](\pbackx,\pbacky)
\global\advance\pbackx by 500
\put(\pbackx,\pbacky){$k$}
\drawline\photon[\E\REG](\Xone,\Yone)[6]
\startphantom
\drawline\fermion[\SW\REG](\pfrontx,\pfronty)[2000]
\stopphantom
\drawline\fermion[\SW\REG](\pbackx,\pbacky)[2000]
\global\advance\pbackx by -1900
\put(\pbackx,\pbacky){$p_-$}
\drawarrow[\NE\ATBASE](\pmidx,\pmidy)
\drawline\fermion[\NE\REG](\photonbackx,\photonbacky)[4000]
\drawarrow[\SW\ATBASE](\pmidx,\pmidy)
\global\advance\pbackx by 500
\put(\pbackx,\pbacky){$q_+$}
\put(\pfrontx,\pfronty){\circle*{500}}
\drawline\fermion[\SE\REG](\photonbackx,\photonbacky)[4000]
\drawarrow[\SE\ATBASE](\pmidx,\pmidy)
\global\advance\pbackx by 500
\put(\pbackx,\pbacky){$q_-$}
\end{picture}
\end{center}

\noindent We can proceed as above and obtain
$${\cal A}^{(ISR)}_{+,+,\lambda_3,\lambda_4,-}=\frac{i\sqrt{2}e^3}{s^\prime}
\times$$
\begin{equation}
\frac{s_{+-}(p_-,p_+)Z_{+,+,\lambda_3,\,\lambda_4}
^{0\, ,\,0\, ,\,1\; ,-1}(p_+,p_-,q_-,q_+)-
s_{+-}(k,p_+)Z_{+,+,\lambda_3,\,\lambda_4}
^{0\, ,\,0\, ,\,1\; ,-1}(p_+,k,q_-,q_+)}
{s_{+-}(k,p_-) s_{+-}(k,p_+)}.
\label{AISRpmm} \end{equation}
We do not give any calculation details because they completely resemble the
ones described for ${\cal A}^{(ISR)}_{+,+,\lambda_3,\lambda_4,+}$.
Maybe only just one little difference deserves to be commented: in calculation
of
$${\bar u}_{\lambda_3}(q_-)\hat \epsilon_-^*(k;p_+)v_{-\lambda_4}(q_+)=
\frac{{\bar u}_-(k)\gamma^\mu u_-(p_+){\bar u}_{\lambda_3}(q_-)\gamma_\mu
v_{-\lambda_4}(q_+)}{\sqrt{2}s_{+-}(k,p_+)}$$
one needs to use the line-reversal identity (\ref{reverseid})
$${\bar u}_-(k)\gamma^\mu u_-(p_+)={\bar u}_+(p_+)\gamma^\mu u_+(k)$$
to obtain
$${\bar u}_{\lambda_3}(q_-)\hat \epsilon_-^*(k;p_+)v_{-\lambda_4}(q_+)=\frac
{Z_{+,+,\lambda_3,\,\lambda_4}^{0\, ,\,0\, ,\,1\; ,-1}
(p_+,k,q_-,q_+)}{\sqrt{2}s_{+-}(k,p_+)}.$$

The remaining helicity amplitudes for initial state radiation are calculated
analogously (or they can be obtained by the parity invariance argument for 
free). The results are
$${\cal A}^{(ISR)}_{-,-,\lambda_3,\lambda_4,+}=\frac{i\sqrt{2}e^3}{s^\prime}
\times$$
\begin{equation}
\frac{s_{-+}(p_-,p_+)Z_{-,-,\lambda_3,\,\lambda_4}
^{0\, ,\,0\, ,\,1\; ,-1}(p_+,p_-,q_-,q_+)-
s_{-+}(k,p_+)Z_{-,-,\lambda_3,\,\lambda_4}
^{0\, ,\,0\, ,\,1\; ,-1}(p_+,k,q_-,q_+)}
{s_{-+}(k,p_-) s_{-+}(k,p_+)}
\label{AISRmpp} \end{equation}
and
$${\cal A}^{(ISR)}_{-,-,\lambda_3,\lambda_4,-}=\frac{i\sqrt{2}e^3}{s^\prime}
\times$$
\begin{equation}
\frac{s_{+-}(p_-,p_+)Z_{-,-,\lambda_3,\,\lambda_4}
^{0\, ,\,0\, ,\,1\; ,-1}(p_+,p_-,q_-,q_+)-
s_{+-}(p_-,k)Z_{-,-,\lambda_3,\,\lambda_4}
^{0\, ,\,0\, ,\,1\; ,-1}(k,p_-,q_-,q_+)}
{s_{+-}(k,p_-) s_{+-}(k,p_+)}.
\label{AISRmpm} \end{equation}

\subsection{Final state radiation}
For the final state radiation both possible diagrams contribute owing to the
nonzero muon mass. So we will not try to get any compact expressions for
${\cal A}^{(FSR)}$, even of the kind we had for ${\cal A}^{(ISR)}$. Instead we
will follow the philosophy of \cite{MMM} and will express the corresponding
amplitudes in terms of the $Z$-functions assuming further numerical evaluation
on a computer.

Let us consider the photon emission from the outgoing muon:
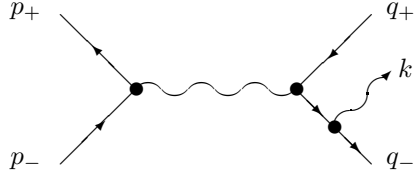
\begin{center}
\begin{picture}(20000,12000)
\drawline\fermion[\SE\REG](2000,9000)[4000]
\global\Xone=\pbackx
\global\Yone=\pbacky
\drawarrow[\NW\ATBASE](\pmidx,\pmidy)
\put(\pbackx,\pbacky){\circle*{500}}
\global\advance\pfrontx by -1900
\put(\pfrontx,\pfronty){$p_+$}
\drawline\fermion[\SW\REG](\pbackx,\pbacky)[4000]
\drawarrow[\NE\ATBASE](\pmidx,\pmidy)
\global\advance\pbackx by -1900
\put(\pbackx,\pbacky){$p_-$}
\drawline\photon[\E\REG](\Xone,\Yone)[6]
\drawline\fermion[\NE\REG](\photonbackx,\photonbacky)[4000]
\drawarrow[\SW\ATBASE](\pmidx,\pmidy)
\global\advance\pbackx by 500
\put(\pbackx,\pbacky){$q_+$}
\put(\pfrontx,\pfronty){\circle*{500}}
\drawline\fermion[\SE\REG](\photonbackx,\photonbacky)[2000]
\drawarrow[\SE\ATBASE](\pmidx,\pmidy)
\put(\pbackx,\pbacky){\circle*{500}}
\drawline\photon[\NE\REG](\pbackx,\pbacky)[3]
\drawarrow[\NE\ATBASE](\pbackx,\pbacky)
\global\advance\pbackx by 500
\put(\pbackx,\pbacky){$k$}
\drawline\fermion[\SE\REG](\photonfrontx,\photonfronty)[2000]
\drawarrow[\SE\ATBASE](\pmidx,\pmidy)
\global\advance\pbackx by 500
\put(\pbackx,\pbacky){$q_-$}
\end{picture}
\end{center}
The corresponding contribution to the ${\cal A}^{(FSR)}_{\lambda_1,\lambda_1,
\lambda_3,\lambda_4,\lambda}$ helicity amplitude has the form
$${\cal A}=\frac{ie^3}{s}~\frac{1}{2q_-\cdot k}{\bar v}_{-\lambda_1}(p_+)
\gamma_\mu u_{\lambda_1}(p_-){\bar u}_{\lambda_3}(q_-)\hat \epsilon^*_\lambda
(k;p)(\hat q_-+m_\mu+\hat k)\gamma^\mu v_{-\lambda_4}(q_+),$$
where $s=(p_-+p_+)^2$. Using 
$$\hat \epsilon^*_\lambda(k;p)=\frac{{\bar u}_\lambda(k)\gamma^\mu u_
\lambda(p)}{\sqrt{2} s_{-\lambda,\lambda}(k,p)}\gamma_\mu$$
and decomposing again ${\cal A}={\cal A}_1+{\cal A}_2$, where ${\cal A}_1$
corresponds to the part originated from
$$\hat q_-+m_\mu=u_+(q_-){\bar u}_+(q_-)+u_-(q_-){\bar u}_-(q_-)$$
and ${\cal A}_2$ -- to the part originated from
$$\hat k=u_+(k){\bar u}_+(k)+u_-(k){\bar u}_-(k),$$
we get
$${\cal A}_1=\frac{ie^3}{s}~\frac{1}{2q_-\cdot k}~\frac{X^{(1)}_++X^{(1)}_-}
{\sqrt{2} ~s_{-\lambda,\lambda}(k,p)}.$$
Here 
$$X^{(1)}_+={\bar v}_{-\lambda_1}(p_+)\gamma_\mu u_{\lambda_1}(p_-)
{\bar u}_{\lambda_3}(q_-){\bar u}_\lambda(k)\gamma^\nu u_\lambda(p)\gamma_\nu
u_+(q_-){\bar u}_+(q_-)\gamma^\mu v_{-\lambda_4}(q_+)=$$
$${\bar v}_{-\lambda_1}(p_+)\gamma_\mu u_{\lambda_1}(p_-)
{\bar u}_+(q_-)\gamma^\mu v_{-\lambda_4}(q_+)
~{\bar u}_\lambda(k)\gamma^\nu u_\lambda(p)
~{\bar u}_{\lambda_3}(q_-)\gamma_\nu u_+(q_-).$$
Therefore
\begin{equation}
X^{(1)}_+=Z_{\lambda_1,\lambda_1,+,\,\lambda_4}^{0\;\, ,\,0\; ,\,1\, ,-1}
(p_+,p_-,q_-,q_+)
*Z_{\lambda,\,\lambda,\lambda_3,+}^{0 ,\,0\, ,\,1\; ,1}(k,p,q_-,q_-).
\label{x1p}\end{equation}
$X^{(1)}_-$ is obtained from $X^{(1)}_+$ simply by reversing the intermediate
helicity sign:
\begin{equation}
X^{(1)}_-=Z_{\lambda_1,\lambda_1,-,\,\lambda_4}
^{0\;\, ,\,0\; ,\,1\;,-1}(p_+,p_-,q_-,q_+)
*Z_{\lambda,\,\lambda,\lambda_3,-}^{0 ,\,0\, ,\,1\; ,1}(k,p,q_-,q_-).
\label{x1m}\end{equation} 
${\cal A}_2$ is dealt with an analogous way as well as contributions from the
diagram with the photon emission from the outgoing $\mu^+$. So we skip the
details and give the final result
$${\cal A}^{(FSR)}_{\lambda_1,\lambda_1,\lambda_3,\lambda_4,\lambda}=
\frac{ie^3}{s}~\frac{1}{2\sqrt{2}~s_{-\lambda,\lambda}(k,p)}\times $$
\begin{equation}
\left \{
\frac{X^{(1)}_++X^{(1)}_-+X^{(2)}_++X^{(2)}_-}{k\cdot q_-}-
\frac{X^{(3)}_++X^{(3)}_-+X^{(4)}_++X^{(4)}_-}{k\cdot q_+}\right \}.
\label{AFSR}\end{equation}
$X^{(1)}_\pm$ were defined above by (\ref{x1p}) and (\ref{x1m}). For the
remaining $X^{(i)}_\pm$ functions the corresponding expressions are given
below:
\begin{eqnarray} &&
X^{(2)}_+=Z_{\lambda_1,\lambda_1,+,\;\lambda_4}^{0\;\,,\,0\;,\,0\;,-1}
(p_+,p_-,k,q_+)
*Z_{\lambda,\,\lambda,\lambda_3,+}^{0,\,0\,,\,1\;,\,0}(k,p,q_-,k)\;,
\nonumber \\ &&
X^{(2)}_-=Z_{\lambda_1,\lambda_1,-,\;\lambda_4}^{0\;\,,\,0\;,\,0\;,-1}
(p_+,p_-,k,q_+)
*Z_{\lambda,\,\lambda,\lambda_3,-}^{0,\,0\,,\,1\;,\,0}(k,p,q_-,k)\;,
\nonumber \\ &&
X^{(3)}_+=Z_{\lambda_1,\lambda_1,\lambda_3,\;+}^{0\;\,,\,0\;,\,1\;\,,-1}
(p_+,p_-,q_-,q_+)
*Z_{\lambda,\,\lambda,\,+\;,\,\lambda_4}^{0,\,0,-1,-1}(k,p,q_+,q_+)\;,
\\ &&
X^{(3)}_-=Z_{\lambda_1,\lambda_1,\lambda_3,\;-}^{0\;\,,\,0\;,\,1\;\,,-1}
(p_+,p_-,q_-,q_+)
*Z_{\lambda,\,\lambda,\,-\;,\,\lambda_4}^{0,\,0,-1,-1}(k,p,q_+,q_+)\;,
\nonumber \\ &&
X^{(4)}_+=Z_{\lambda_1,\lambda_1,\lambda_3,+}
^{0\;\,,\,0\;,\,1\;\,,\,0}(p_+,p_-,q_-,k,)
*Z_{\lambda,\,\lambda,\,+,\,\lambda_4}^{0,\,0,\,0\;,-1}(k,p,k,q_+)\;,
\nonumber \\ &&
X^{(4)}_-=Z_{\lambda_1,\lambda_1,\lambda_3,-}
^{0\;\,,\,0\;,\,1\;\,,\,0}(p_+,p_-,q_-,k,)
*Z_{\lambda,\,\lambda,\,-,\,\lambda_4}^{0,\,0,\,0\;,-1}(k,p,k,q_+)\;.
\nonumber \label{x24pm}\end{eqnarray}
To facilitate the massless limit, we make the following helicity-dependent 
choice for the auxiliary four-vector $p$:
\begin{equation}
p=\left \{ \begin{array}{c} q_{+\xi}, \;\; {\mathrm if} \;\; 
\lambda\lambda_4=+1,
\\ q_{-\xi}, \;\; {\mathrm if} \;\; 
\lambda\lambda_4=-1. \end{array} \right .
\label{pdef} \end{equation}

\subsection{Squared matrix element in the massless limit}
Let us check our helicity amplitudes against $m_\mu\to 0$ massless limit. 
For illustration we will sketch the derivation of the limit for 
${\cal A}_{+,+,-,-,+}$. In the massless limit we have $s^\prime=2q_-\cdot
q_+=-s_{+-}(q_+,q_-)s_{-+}(q_+,q_-)$ and
$$Z(+,+,-,-)=-2s_{+-}(p_1,p_4)s_{-+}(p_2,p_3).$$
Therefore (\ref{AISRpmp}) takes the form
$${\cal A}_{+,+,-,-,+}^{(ISR)}=2\sqrt{2}ie^3 \times $$ $$
\frac{s_{-+}(p_-,q_-)\left [
s_{-+}(p_-,p_+)s_{+-}(p_+,q_+)-s_{-+}(p_-,k)s_{+-}(k,q_+)\right ]}
{s_{+-}(q_+,q_-)s_{-+}(q_+,q_-)s_{-+}(k,p_-)s_{-+}(k,p_+)}. $$
But using four-momentum conservation $p_-+p_+=q_-+q_++k$ and the Dirac 
equation we get
$$\left [s_{-+}(p_-,p_+)s_{+-}(p_+,q_+)-s_{-+}(p_-,k)s_{+-}(k,q_+)\right ]= $$
$${\bar u}_-(p_-)\left [ u_+(p_+){\bar u}_+(p_+)-u_+(k){\bar u}_+(k)\right ]
u_-(q_+)=$$ $${\bar u}_-(p_-)\omega_+(\hat p_+-\hat k)u_-(q_+)=
{\bar u}_-(p_-)\omega_+(\hat q_-+\hat q_+-\hat p_-)u_-(q_+)= $$ $$
{\bar u}_-(p_-)u_+(q_-){\bar u}_+(q_-)u_-(q_+)=s_{-+}(p_-,q_-)
s_{+-}(q_-,q_+),$$
and ${\cal A}_{+,+,-,-,+}^{(ISR)}$ is simplified to
$${\cal A}_{+,+,-,-,+}^{(ISR)}=-2\sqrt{2}ie^3\frac{s^2_{-+}(p_-,q_-)}{
s_{-+}(q_+,q_-)s_{-+}(k,p_-)s_{-+}(k,p_+)}= $$
\begin{equation}
2ie^3\frac{s^2_{-+}(p_-,q_-)}{s_{-+}(p_-,p_+)s_{-+}(q_-,q_+)}
~\beta_+(p_-,p_+,k),
\label{ISR0} \end{equation}
Where $\beta_+(p_+,p_-,k)$ is defined by (\ref{beta}).
For the  given helicity configuration, in the massless limit, $p=q_-$ and all
$X^{(i)}_+$ functions in (\ref{AFSR}) equal zero. $X^{(1)}_-$ and $X^{(2)}_-$
are proportional to $s_{-+}(q_-,q_-)$ and hence also are zeros. The only
nonzero combinations are
$$X^{(3)}_-=4s_{+-}(p_+,q_+)s_{-+}(p_-,q_-)s_{+-}(k,q_+)s_{-+}(q_-,q_+)$$
and
$$X^{(4)}_-=4s_{+-}(p_+,k)s_{-+}(p_-,q_-)s_{+-}(k,q_+)s_{-+}(q_-,k).$$
So (\ref{AFSR}) is reduced to
$${\cal A}_{+,+,-,-,+}^{(FSR)}=-2\sqrt{2}ie^3\times$$
$$\frac{s_{-+}(p_-,q_-)s_{+-}(k,q_+)\left [s_{+-}(p_+,q_+)
s_{-+}(q_-,q_+)+ s_{+-}(p_+,k)s_{-+}(q_-,k)\right ]}{s~2k\cdot q_+~s_{-+}
(k,q_-)}.$$
We can apply four-momentum conservation and the Dirac equation again to 
simplify
$$\left [s_{+-}(p_+,q_+)s_{-+}(q_-,q_+)+ s_{+-}(p_+,k)s_{-+}(q_-,k)\right ]=
s_{+-}(p_-,p_+)s_{-+}(p_-,q_-).$$
Besides
$$s=2p_-\cdot p_+=-s_{+-}(p_-,p_+)s_{-+}(p_-,p_+),\;\;
2k\cdot q_+=-s_{+-}(k,q_+)s_{-+}(k,q_+).$$
Therefore
$${\cal A}_{+,+,-,-,+}^{(FSR)}=-2\sqrt{2}ie^3\frac{s^2_{-+}(p_-,q_-)}
{s_{-+}(p_-,p_+)s_{-+}(k,q_-)s_{-+}(k,q_+)}=$$ 
\begin{equation}
-2ie^3\frac{s^2_{-+}(p_-,q_-)}{s_{-+}(p_-,p_+)s_{-+}(q_-,q_+)}
\beta_+(q_-,q_+,k).
\label{FSR0} \end{equation}
Combining (\ref{ISR0}) and (\ref{FSR0}) we get in the massless limit
\begin{equation}
{\cal A}_{+,+,-,-,+}=2ie^3\frac{s^2_{-+}(p_-,q_-)}{s_{-+}(p_-,p_+)
s_{-+}(q_-,q_+)}\left [\beta_+(p_-,p_+,k)-\beta_+(q_-,q_+,k)\right ].
\label{Appmmp0} \end{equation}
This result agrees to the one given in \cite{Xu} (note the sign difference
in our definition of the photon polarization four-vector (\ref{epsilon}) 
compared to what is used in \cite{Xu}).

The massless limit for other helicity configurations can be considered 
similarly and the results are given below:
\begin{eqnarray} 
{\cal A}_{+,+,-,-,-}&=&2ie^3\frac{s^2_{-+}(p_+,q_+)}{s_{-+}(p_-,p_+)
s_{-+}(q_-,q_+)}\left [\beta_-(p_-,p_+,k)-\beta_-(q_-,q_+,k)\right ],
\nonumber \\ 
{\cal A}_{+,+,+,+,+}&=&-2ie^3\frac{s^2_{-+}(p_-,q_+)}{s_{-+}(p_-,p_+)
s_{-+}(q_-,q_+)}\left [\beta_+(p_-,p_+,k)-\beta_+(q_-,q_+,k)\right ],
\nonumber \\ 
{\cal A}_{+,+,+,+,-}&=&-2ie^3\frac{s^2_{-+}(p_+,q_-)}{s_{-+}(p_-,p_+)
s_{-+}(q_-,q_+)}\left [\beta_-(p_-,p_+,k)-\beta_-(q_-,q_+,k)\right ], 
\nonumber \\
\label{A0}\end{eqnarray}
The amplitudes for the remaining helicity configurations are obtained
by simply reversing the signs of the helicity labels in the above given
expressions (parity conjugation).

To evaluate the norms of the above given massless amplitudes, we will need
$|\beta_\lambda(p_-,p_+,k)-\beta_\lambda(q_-,q_+,k)|^2$. Note that
\begin{equation}
\beta_\lambda(p_-,p_+,k)=\frac{p_+\cdot \epsilon_\lambda (k,p_-)}
{k\cdot p_+}=\frac{{\bar u}_\lambda (k)\hat p_+ u_\lambda(p_-)}
{\sqrt{2}k\cdot p_+ s_{-\lambda,\lambda}(k,p_-)}.
\label{beta_lambda}
\end{equation}
Let us introduce a four-vector \cite{Berends}\footnote{To motivate its 
introduction, note that the first term is already present in 
(\ref{beta_lambda}), and we can add any term proportional to $p_-$ without
changing the result, since $\hat p_-u_\lambda(p_-)=0$. The proportionality 
coefficient is chosen so that $\hat k $ and $\hat v_p $ commute, as this will 
allow the calculation to be performed as described in the text.} 
\begin{equation}
v_p=\frac{p_+}{k\cdot p_+}-\frac{p_-}{k\cdot p_-},
\label{vp} \end{equation}
such that $k\cdot v_p=0$ and therefore $\hat k \hat v_p=- \hat v_p \hat k$.
Then
$$ \beta_\lambda(p_-,p_+,k)=\frac{{\bar u}_\lambda (k)\hat v_p u_\lambda(p_-)}
{\sqrt{2}s_{-\lambda,\lambda}(k,p_-)}.$$
But 
$${\bar u}_\lambda (k)\hat v_p u_\lambda(p_-)=\frac{{\bar u}_{-\lambda} (\xi)
\hat k \hat v_p u_\lambda (p_-)}{\sqrt{2k\cdot \xi}}=
-\frac{{\bar u}_{-\lambda} (\xi)\hat v_p\hat k
u_\lambda (p_-)}{\sqrt{2k\cdot \xi}}.$$
Substituting here $\hat k=u_\lambda (k) {\bar u}_\lambda (k)+
u_{-\lambda} (k) {\bar u}_{-\lambda} (k)$ we get
$${\bar u}_\lambda (k)\hat v_p u_\lambda(p_-)=-s_{-\lambda,\lambda}(k,p_-)~
\frac{{\bar u}_{-\lambda} (\xi)\hat v_p u_{-\lambda}(k)}{\sqrt{2k\cdot \xi}}$$
and therefore
\begin{equation}
\beta_\lambda(p_-,p_+,k)=-\frac{{\bar u}_{-\lambda} (\xi)\hat v_p 
u_{-\lambda}(k)}{2\sqrt{k\cdot \xi}}.
\label{betap} \end{equation}
Then it is easy to get
$$\left |\beta_\lambda(p_-,p_+,k)\right |^2=\frac{Sp\left \{\omega_{-\lambda}
\hat \xi \hat v_p \hat k \hat v_p\right \}}{4k\cdot \xi}.$$
But $\hat v_p \hat k \hat v_p=-v_p^2\hat k$ and therefore we get
$$\left |\beta_\lambda(p_-,p_+,k)\right |^2=-\frac{1}{2}v_p^2.$$
Analogously
$$\beta_\lambda(p_-,p_+,k)\beta^*_\lambda(q_-,q_+,k)+
\beta_\lambda(q_-,q_+,k)\beta^*_\lambda(p_-,p_+,k)=$$ $$
\frac{Sp\left \{\omega_{-\lambda}\hat \xi
\left (\hat v_p \hat k \hat v_q
+\hat v_q \hat k \hat v_p\right ) \right \} }{4k\cdot \xi}=-v_p\cdot v_q. $$
Therefore
\begin{equation}
\left |\beta_\lambda(p_-,p_+,k)-\beta_\lambda(q_-,q_+,k)\right |^2=
-\frac{1}{2}\left (v_p-v_q\right )^2.
\label{betapq} \end{equation}
This equation enables to find the massless limit for the squared matrix 
element averaged over initial and summed over final polarizations as
\begin{equation}
\overline {|{\cal A}|^2}=-e^6~\frac{t^2+u^2+t^{\prime 2}+u^{\prime 2}}
{s s^\prime}\left (v_p-v_q\right )^2,
\label{A20} \end{equation}
where
\begin{eqnarray} &&
s=(p_-+p_+)^2\;,\;\;
t=-2p_+\cdot q_+,\;\;
u=-2p_+\cdot q_-, \nonumber \\ &&
s^\prime=(q_-+q_+)^2,\;\;
t^\prime=-2p_-\cdot q_-,\;\;
u^\prime=-2p_-\cdot q_+.
\label{invariants} \end{eqnarray}
This result also agrees to what is known in the literature \cite{Berends,Xu}
and can be used in the program as a testing tool for the tree level amplitude
against bugs.

As an another testing tool we can use the hard bremsstrahlung squared 
amplitude from \cite{BKJW} which incorporates the nonzero muon mass. In our
notations and for the massless electron  this result is
\begin{equation}
\overline {|{\cal A}|^2}=e^6(R_{ini}+R_{fin}+R_{int}),
\label{A2BKJW} \end{equation}
where individual contributions, from the initial state radiation, final state
radiation and their interference, are given by
\begin{eqnarray} &&
R_{ini}=\frac{1}{s^\prime x_1x_2}\left \{ t^2+u^2+t^{\prime 2}+u^{\prime 2}+
2m_\mu^2~\frac{(t+u)^2+(t^\prime+u^\prime)^2}{s^\prime}\right \},
\nonumber \\ &&
R_{fin}=\frac{1}{s y_1 y_2}\left \{ \left [\frac{}{}
 t^2+u^{\prime 2}+2m_\mu^2s\right ]
\left [ 1-2\frac{m_\mu^2}{s}\left ( 1+\frac{y_1}{y_2}\right ) \right ]+
\nonumber \right . \\ && \left . 
\hspace*{-5mm}\left [\frac{}{} t^{\prime 2}+u^2+2m_\mu^2s\right ]
\left [ 1-2\frac{m_\mu^2}{s}\left ( 1+\frac{y_2}{y_1}\right ) \right ]+
8\frac{m_\mu^2}{s}\left (x_1^2+x_2^2 \right )-8m_\mu^2(s-s^\prime)\right \},
\nonumber \\ && 
R_{int}=\left . \left . \frac{1}{ss^\prime x_1x_2y_1y_2} 
\right \{ \left (\frac{}{} ux_2y_1+u^\prime
x_1y_2-tx_2y_2-t^\prime x_1y_1\right ) \times \right . \label{Rini}
\\ && \left .
\hspace*{38mm} \left (\frac{}{} t^2+t^{\prime 2}+y^2+u^{\prime 2}
+2m_\mu^2(s+s^\prime) \right )+ \right . \nonumber \\ && \left .
+2m_\mu^2x_1x_2\left [\frac{}{} (s-s^\prime)(u+u^\prime-t-t^\prime)-
4(x_1-x_2)(y_1-y_2)\frac{}{} \right ]\right \}, \nonumber 
\end{eqnarray}
with
\begin{equation}
x_1=k\cdot p_+,\;\; x_2=k\cdot p_-,\;\; y_1=k\cdot q_+,\;\;
y_2=k\cdot q_-\;. \label{xydef}
\end{equation}   

The same squared amplitude was calculated in \cite{AFKM}. Although the formulas
look different, we have checked numerically in the massless electron limit 
that both \cite{AFKM} and \cite{BKJW} give the same result (after the 
$\Delta_{s_1s_1}$ sign misprint had been corrected in formula (2.12) of
\cite{AFKM}). Our helicity amplitudes presented above also have been checked
numerically to lead to the same tree level 
$e^-e^+\to\mu^-\mu^+\gamma$ squared amplitude (\ref{A2BKJW}-\ref{Rini}). 
  
\section{Monte Carlo algorithm}
The matrix element we have described in the previous section is strongly 
peaked in certain regions of the phase space. For this reason generating 
events distributed according to this matrix element efficiently is not 
a trivial task. In this section a detailed description of the Monte Carlo 
algorithm used in the event generator is given.

In the next subsection, we will start with some general remarks on how random 
numbers with a given probability distribution can be generated. In particular,
we describe the acceptance-rejection algorithm that will be used in our Monte
Carlo program. We then explain in seperate subsections how the initial and 
final state radiations can be described by crude distributions that can be 
efficiently generated using combinations of analytic inversion and 
acceptance-rejection methods.

The separation into an initial-state crude distribution and the final-state 
crude distribution means that we are not considering the interference between
them. This is done to facilitate the Monte Carlo generation of these 
distributions. The interference is automatically reintroduced in the last
step of the acceptance-rejection algorithm because we use the ratio of the 
square of the actual amplitude to the sum of the squares of the initial-state 
and final-state crude amplitudes to calculate the acceptance-rejection weight.

For the acceptance-rejection method to be effective, it is important that the 
coarse distribution reproduces the main features of the real distribution. In
our case, the physical origin of the various singularities of the distribution 
to be generated is that virtual fermions in the Feynman diagrams describing 
radiation from the initial and final states  may appear near on-shell in some 
kinematic situations. For radiation in the initial state from a massless (in 
our approximation) electron or positron, this occurs for the collinear 
radiation or when a very soft photon is emitted. For radiation from a muon or 
anti-muon in the final state, the peaked region in the amplitude corresponds 
to the emission of soft photons. 
  
\subsection{General remarks}
The most efficient way to generate random numbers with a given probability
density $f(x)$ is as follows. If we define
\begin{equation}
r=\int\limits_{x_{min}}^x f(y)dy\equiv F(x),
\label{ainv}\end{equation}
then the probability that in a given event $x$-variable lays in the interval 
$[x,x+dx]$ is $f(x)dx=dr$. However, this is also the probability that
the variable $r$
lays in the interval $[r,r+dr]$, where $r=F(x)$. Therefore $r$-variable
is distributed uniformly. This gives the following algorithm to generate 
$f(x)$ distribution (which will be assumed to be normalized in the interval 
$[x_{min},x_{max}]$, although this is not too relevant):
\begin{itemize}
\item generate a random number $r$ uniformly distributed in the interval
$[0,1]$.
\item find $x$ such that $r=F(x)$. 
\end{itemize}
This is the so called analytic inversion method \cite{Wnzrl}. It is assumed 
that (\ref{ainv}) gives an analytic expression for $F(x)$ which can be easily
inverted, at least numerically. Unfortunately this is not always the case.
In such situations acceptance-rejection method turns out to be helpful. 
Suppose
$f_0(x)$ is some crude distribution which: 1) shares major features of
desired $f(x)$ distribution and 2) can be generated by analytic 
inversion. Then the acceptance-rejection algorithm to generate 
$f(x)$-distribution goes as follows \cite{Wnzrl,PGMC}:
\begin{itemize}
\item generate a random number $x$ according to the crude distribution 
$f_0(x)$.
\item calculate the weight $w=f(x)/f_0(x)$.
\item generate a random number $r$ uniformly distributed in the interval
$[0,C]$, where $C$ is some number larger than maximal weight $w_{max}$ (not
far from it, however, if a good efficiency is desired).
\item if $r\le w$, accept the event, otherwise repeat the whole procedure.
\end{itemize}
The probability that the first step will produce $x$-variable in the interval
$[x,x+dx]$ is $p_1=f_0(x)dx$, while the probability of accepting this event
is $p_2=w/C$. Therefore, the combined probability is
$$p=p_1p_2=\frac{1}{C}f_0(x)wdx=\frac{1}{C}f(x)dx$$
and the accepted events are indeed distributed according to the density 
function $f(x)$.

In our case we want to generate the distribution
\begin{equation}
d\sigma=\frac{\alpha^3}{8\pi^2s}~R~\delta(Q-q_--q_+)
~\frac{d\vec{q}_-}{E_-}~\frac{d\vec{q}_+}{E_+}~\frac{d\vec{k}}{\omega}\;,
\label{dsigma} \end{equation}
where $E_\pm,\;\omega$ stand for the $\mu^\pm$ and photon 
energies, $Q=p_-+p_+-k$,
\begin{equation}
R=\frac{\overline{|{\cal{A}}|^2}}{e^6},
\label{Rdef} \end{equation}
and the matrix element $\overline{|{\cal{A}}|^2}$ was discussed earlier. The 
crude distribution, that models peculiar features of $R$ consists of two 
parts
\begin{equation}
d\sigma_0=d\sigma^{(ISR)}_0+d\sigma^{(FSR)}_0,
\label{dsigma0} \end{equation}
where the first one deals with initial state radiation and the
second one with the final state radiation. The decision which part of the
crude distribution to generate in each event is based on relative
magnitudes of the corresponding total cross sections $\sigma^{(ISR)}_0$ and
$\sigma^{(FSR)}_0$. After event is generated according to the crude 
distribution, the acceptance-rejection method is applied to restore the real
distribution.

Given the total number of events $N_{tot}$, generated with 
the crude distribution, and the number of accepted events $N_{acc}$, the
total cross section can be calculated as
\begin{equation}
\sigma(e^+e^-\to\mu^+\mu^-\gamma)=\frac{N_{acc}}{N_{tot}}\left (
\sigma^{(ISR)}_0+\sigma^{(FSR)}_0\right )C,
\label{totsigma}\end{equation}
where $C$ is the majoring value for the weights,
since the acceptance probability of an average event is given by
\begin{equation}
p=\frac{1}{C}~\frac{\sigma(e^+e^-\to\mu^+\mu^-\gamma)}
{\sigma^{(ISR)}_0+\sigma^{(FSR)}_0},
\label{avacp}\end{equation}
and, on the other hand, $p=N_{acc}/N_{tot}$ for sufficiently large $N_{tot}$.
Since $N_{acc}$ is distributed according to the 
binomial distribution, for an error estimate one can take
\begin{equation}
\Delta\sigma=\sqrt{\frac{p(1-p)}{N_{tot}}}\left (
\sigma^{(ISR)}_0+\sigma^{(FSR)}_0\right )C.
\label{ertsigma}\end{equation}
  
Now we describe in some details how $d\sigma^{(ISR)}_0$ and 
$d\sigma^{(FSR)}_0$ distributions are generated. Our treatment was inspired
by \cite{BKJets,BKmmg,CPC29}, but explicit realization of the 
algorithms is original.
 
\subsection{Crude distribution for the initial state radiation}
To approximate initial state radiation part of
$e^-e^+\to\mu^-\mu^+\gamma$ matrix element, we use 
\begin{equation}
R^{(ISR)}_0=
\frac{1}{s^\prime x_1 x_2 \beta^*}
\left \{ (t+u)^2+(t^\prime+u^\prime)^2 \right \},
\label{RISR} \end{equation}
where
$$\beta^*=\sqrt{1-\frac{4m_\mu^2}{s^\prime}}$$
is the muon velocity in the dimuon rest frame and
other invariants used in (\ref{RISR}) were defined earlier in 
(\ref{invariants}) and (\ref{xydef}). 

After some trial and error, the form (\ref{RISR}) was motivated by the 
second term in the $R_{ini}$ expression in (\ref{Rini}). The factor
$1/\beta^*$ was added to eliminate it after integration over the
muon and antimuon momenta (see (\ref{betastarint}).

Let us explain how $d\sigma^{(ISR)}_0$ distribution is generated. First of
all we need the corresponding energy and angular spectra of the photon. To get
those, we integrate the distribution
$$d\sigma^{(ISR)}_0=\frac{\alpha^3}{8\pi^2s}~R^{(ISR)}_0~\delta(Q-q_--q_+)
~\frac{d\vec{q}_-}{E_-}~\frac{d\vec{q}_+}{E_+}~\frac{d\vec{k}}{\omega}$$
over the muon and antimuon momenta. But since
$$(t+u)^2+(t^\prime+u^\prime)^ 2=4\left [ (p_-\cdot Q)^2+(p_+\cdot Q)^2
\right ], $$
the integrand does not depend on muon and antimuon momenta.
Using
\begin{equation}
\int \delta(Q-q_--q_+)~\frac{d\vec{q}_-}{E_-}~\frac{d\vec{q}_+}{E_+}=
2\pi\beta^* ,
\label{betastarint}
\end{equation}
we obtain after integrating over those momenta
$$d\sigma^{(ISR)}_0=\frac{\alpha^3}{\pi}~\frac{1}{ss^\prime x_1x_2} \left [
(p_+\cdot Q)^2+(p_-\cdot Q)^2\right ]~\frac{d\vec{k}}{\omega}.$$
Using
$$(p_+\cdot Q)^2+(p_-\cdot Q)^2=\frac{1}{2}\left\{ \left [(p_++p_-)\cdot Q
\right ]^2+\left [(p_+-p_-)\cdot Q \right ]^2 \right\}=$$
$$= \frac{1}{2}\left\{ [s-(x_1+x_2)]^2+(x_1-x_2)^2\right \}=
\frac{1}{2}\left\{ s^2-2s(x_1+x_2)+2(x_1^2+x_2^2)\right \}$$
and substituting 
$$x_1+x_2=\frac{s-s^\prime}{2},\;\;
x_1^2+x_2^2=\frac{(s-s^\prime)^2}{4}-2x_1x_2,$$
we obtain
$$(p_+\cdot Q)^2+(p_-\cdot Q)^2=2\left [\frac{s^2+s^{\prime 2}}{8}-x_1x_2
\right ].$$
Therefore
\begin{eqnarray} &&
d\sigma^{(ISR)}_0=\frac{2\alpha^3}{\pi}~\frac{1}{ss^\prime}\left [
\frac{s^2+s^{\prime 2}}{8x_1x_2}-1\right ]~\frac{d\vec{k}}{\omega}=
\nonumber \\ &&
=\frac{4\alpha^3}{ss^\prime}\left [ \frac{s^2+s^{\prime 2}}
{2s\omega^2(1-\cos^2{\theta})}-1\right ]\omega d\omega ~d\cos{\theta},
\label{ISRdk} \end{eqnarray}
where $\theta$ stands for the photon polar angle. 

Performing the angular integration in (\ref{ISRdk}) we get the photon energy
spectrum
\begin{equation}
d\sigma^{(ISR)}_0=\frac{2\alpha^3}{s}~\frac{1}{1-x}\left \{2\left (1-x+
\frac{x^2}{2}\right )\ln{\frac{1+c_m}{1-c_m}}-c_mx^2\right\}~\frac{dx}{x},
\label{ISReg} \end{equation}
where the photon energy fraction $x=\omega/E$ and $E$ is the CMS beam energy.
Also, $c_m=\cos{\,\theta_{min}}$ represents the angular cut on the photon. 
Note that
$$1+\frac{s^{\prime 2}}{s^2}=2-2x+x^2.$$  

The distribution (\ref{ISReg}) is generated as follows. As the  first step, 
preliminary distribution (the norm is not included)
$$p_0(x)=\frac{1}{x(1-x)}$$
is generated by analytic inversion. Then the acceptance-rejection
method is applied with the weight
$$w=2\left (1-x+\frac{x^2}{2}\right )\ln{\frac{1+c_m}{1-c_m}}-c_mx^2.$$
The photon energy fraction $x$ is generated in this way in the interval
$x_{min}\le x\le x_{max}$, where
$$x_{max}=1-\frac{m_\mu^2}{E^2}.$$

Now we need to generate angular variables for the photon. The matrix element 
does not depend on the photon azimuthal angle. So this angle $0\le\phi\le 
2\pi$ is generated as an uniform distribution. For the polar angle, 
Eq. (\ref{ISRdk}) shows that we need to generate the distribution
$$f(\cos{\theta})=\frac{1}{1-\cos^2{\theta}}-\frac{x^2}{2(2-2x+x^2)}.$$
Again, we use the analytic inversion method to generate the first term of this 
distribution, which is peaked at small angles, and then acceptance-rejection 
is applied with weights
$$ 1-\frac{x^2(1-\cos^2{\theta})}{2(2-2x+x^2)}.$$
The photon polar angle is restricted to the interval $-c_m\le\cos{\theta}
\le c_m$.

To generate the muon momentum, let us note that
$$\int \delta(Q-q_--q_+)~\frac{d\vec{q}_-}{E_-}~\frac{d\vec{q}_+}{E_+}=
\frac{\beta^*}{2}\int d\cos{\theta_-^*}~d\phi_-^*,$$
where $\theta_-^*$ and $\phi_-^*$ are $\mu^-$ angular variables in the dimuon
rest frame. As $R^{(ISR)}_0$ does not depend on these variables, 
$\cos{\theta_-^*}$ and $\phi_-^*$ are generated as uniform distributions and
then $\mu^-$ four-momentum can be constructed simply in the dimuon rest frame.
Finally, we apply Lorentz transformation to transform this four-momentum back
to the CMS frame.

The total crude ISR cross section is found by integrating (\ref{ISReg}):
$$\sigma_0^{(ISR)}=
=\frac{2\alpha^3}{s}\left [ 2\ln{\frac{x_{max}}{x_{min}}}
\ln{\frac{1+c_m}{1-c_m}}+ \right. $$
\begin{equation}
\left . +\left (\ln{\frac{1+c_m}{1-c_m}}-c_m\right )\left (
\ln{\frac{1-x_{min}}{1-x_{max}}}-x_{max}+x_{min}\right ) \right ].
\label{sg0ISR} \end{equation}

\subsection{Crude distribution for the final state radiation}
As a crude distribution which models peculiarities of the final state 
radiation according to (\ref{Rini}), we take (the role of $s$ is to obtain the 
correct dimension):
\begin{equation}
R_0^{(FSR)}=\frac{s}{y_1y_2}.
\label{RFSR}\end{equation}
Note that
\begin{equation}
y_1+y_2=k\cdot (p_++p_-)=2E(2E-E_+-E_-)
\label{y1plusy2} \end{equation}
and
$$y_1-y_2=(q_+-q_-)\cdot (q_++q_-+k)=2E(E_+-E_-).$$
These two relations enable us to find
\begin{equation}
y_1=2E(2E-E_-),\;\; y_2=2E(2E-E_+).
\label{y1y2}\end{equation}
Now let us transform the phase space 
$$\int  \delta(Q-q_--q_+)
~\frac{d\vec{q}_-}{E_-}~\frac{d\vec{q}_+}{E_+}~\frac{d\vec{k}}{\omega}=
\int d\omega ~d\phi_+^\gamma~\frac{dy_2}{2E}~d\cos{\theta_+}~d\phi_+,$$
where $\theta_+,\;\phi_+$ are $\mu^+$ angular variables and the photon
angles $\theta_+^\gamma,\;\phi_+^\gamma$ are defined with respect to the
$\mu^+$ momentum $\vec{q}_+$ (the photon polar angle $\theta_+^\gamma$ does
not appear in the r.h.s. of the above formula because energy $\delta$-function
enables to integrate over $d\cos{\theta_+^\gamma}$). Therefore the crude 
distribution for final state radiation takes the form
$$d\sigma_0^{(FSR)}=\frac{\alpha^3}{8\pi^2}~\frac{1}{2Ey_1y_2}
d\omega ~d\phi_+^\gamma~dy_2~d\cos{\theta_+}~d\phi_+.$$
It is clear that $\phi_+^\gamma,\;\phi_+$ and $\cos{\theta_+}$ are 
distributed uniformly. Integrating over them and using
$$\frac{1}{y_1y_2}=\frac{1}{2E\omega}\left (\frac{1}{y_1}+\frac{1}{y_2}
\right ),$$
we get $d\sigma_0^{(FSR)}$ as a sum of two distributions
\begin{equation}
d\sigma_0^{(FSR)}=\frac{\alpha^3}{s}~\frac{d\omega}{\omega}~\frac{dy_1}{y_1}+
\frac{\alpha^3}{s}~\frac{d\omega}{\omega}~\frac{dy_2}{y_2},
\label{fsrdwdy} \end{equation}
where we have also used $|dy_2|=|dy_1|$ for fixed $\omega$ as follows from
Eq. (\ref{y1plusy2}).
Integration limits on $y_1$ and $y_2$ are determined in the following way.
From
$$2E-\omega-E_+=E_-=\sqrt{\omega^2+E_+^2+2\omega|\vec{q}_+|\cos{\theta_+
^\gamma}}$$
we find
\begin{equation}
\cos{\theta_+^\gamma}=\frac{2E(E-\omega)-E_+(2E-\omega)}{\omega\sqrt{E_+^2-
m_\mu^2}}.
\label{thpg} \end{equation}
Since $|\cos{\theta_+^\gamma}|\le 1$, we find allowed region for $E_+$:
$$\frac{1}{2}\left [ 2-x-x\sqrt{1-\frac{\mu^2}{1-x}}~\right ]\le x_+\le
\frac{1}{2}\left [ 2-x+x\sqrt{1-\frac{\mu^2}{1-x}}~\right ],$$
where $x_+=E_+/E$ and $\mu=m_\mu/E$. Using now $y_2=2E^2(1-x_+)$, we derive
\begin{equation}
x\left [ 1-\sqrt{1-\frac{\mu^2}{1-x}}\right ]\le\frac{y_2}{E^2}\le
x\left [ 1+\sqrt{1-\frac{\mu^2}{1-x}}\right ].
\label{y2lim}\end{equation}
The limits for $y_1$ are the same as $y_1=2E^2x-y_2$ relation shows. 
Therefore, integrating (\ref{fsrdwdy}) over $y_{1,2}$, we get the following
photon energy spectrum
\begin{equation}
d\sigma_0^{(FSR)}=\frac{2\alpha^3}{s}~\ln{\frac{\sqrt{1-x}+\sqrt{1-x-\mu^2}}
{\sqrt{1-x}-\sqrt{1-x-\mu^2}}}~\frac{dx}{x}.
\label{fsrdw}\end{equation}
Using this distribution, we can generate the photon energy in the following 
way. Noting that 
$$\ln{\frac{\sqrt{1-x}+\sqrt{1-x-\mu^2}}{\sqrt{1-x}-\sqrt{1-x-\mu^2}}}
~\longrightarrow~\frac{\ln{(1-x)}}{x}+\frac{\ln{(s/m_\mu^2)}}{x},$$
when $\mu \to 0$, we first generate the distribution (the norm is not 
included)
$$p_0(x)=\frac{\ln{(1-x)}}{x}+\frac{\ln{(s/m_\mu^2)}}{x}$$
by numerically solving the equation
$$r=\ln{\left(\frac{s}{m_\mu^2}\right )}\ln{\left(\frac{x}{x_{min}}\right )}
+{\mathrm{Li}}_2(x_{min})-{\mathrm{Li}}_2(x),$$
where $r$ is a random number uniformly distributed in the interval
$$0\le r\le\ln{\left(\frac{s}{m_\mu^2}\right )}\ln{\left(\frac{x_{max}}
{x_{min}}\right )}+{\mathrm{Li}}_2(x_{min})-{\mathrm{Li}}_2(x_{max}).$$
Then the desired distribution (\ref{fsrdw}) is reproduced by the 
acceptance-rejection me\-thod with the weight
$$w=\ln{\frac{2(1-x)-\mu^2+2\sqrt{(1-x)(1-x-\mu^2)}}{\mu^2}}\left /
\ln{\frac{4(1-x)}{\mu^2}} \right .  .$$

After the photon energy is generated, we can generate $\mu^+$-energy as well.
First $1/y$ distribution is generated by analytic inversion. Then we 
generate a random number $r$ uniformly distributed in the interval $0\le r 
\le 1$ and if $r\le 0.5$ we take $y_1=y$, otherwise we take $y_2=y$, which
means that 
we choose between two sub-distributions in (\ref{fsrdwdy}). Knowing $y_1$ or
$y_2$ allows us to determine $E_+$ from (\ref{y1y2}) and energy conservation.

As we mentioned earlier, $\phi_+^\gamma,\;\phi_+$ and $\cos{\theta_+}$ are 
generated as uniform distributions and $\cos{\theta^\gamma_+}$ is determined
by (\ref{thpg}). The only thing which is left is to rotate
generated three-momentum of the photon correctly, because 
$\cos{\theta^\gamma_+}$
and $\phi_+^\gamma$ determine its orientation with respect to  
$\mu^+$-momentum $\vec{q}_+$, but not with respect to our initial CMS axes.  
Note that this $q_+$-reference frame can be obtained from the initial 
$p_-$-reference frame (in which $\vec{p}_-$ points along $z$-direction) by 
two consecutive rotations: rotation by the angle $\phi_+$ around $z$-th axis
followed by rotation around the new $y$ axis by the angle $\theta_+$. This 
allows us to construct the transformation law for any vector $\vec{A}=\{ A_x,
A_y,A_z\}$ under the combined rotation:
\begin{eqnarray} &&
A_x=\cos{\theta_+}\cos{\phi_+}A^\prime_x-\sin{\phi_+}A^\prime_y+
\sin{\theta_+}\cos{\phi_+}A^\prime_z, \nonumber \\ &&
A_y=\cos{\theta_+}\sin{\phi_+}A^\prime_x+\cos{\phi_+}A^\prime_y+
\sin{\theta_+}\sin{\phi_+}A^\prime_z, \nonumber \\ &&
A_z=-\sin{\theta_+}A^\prime_x+\cos{\theta_+}A^\prime_z,
\label{rotation}\end{eqnarray}
where $\{ A^\prime_x,A^\prime_y,A^\prime_z\}$ and $\{ A_x,A_y,A_z\}$ are the 
vector coordinates in the $q_+$-reference frame and the desired 
coordinates in the $p_-$-reference frame, respectively. All what is needed 
to complete the generation of the photon momentum is to apply (\ref{rotation}) 
to the case
$\vec{A}=\vec{k}$ and use
$$k^\prime_x=\omega\sin{\theta_+^\gamma}\cos{\phi_+^\gamma},\;\;
k^\prime_y=\omega\sin{\theta_+^\gamma}\sin{\phi_+^\gamma},\;\;
k^\prime_z=\omega\cos{\theta_+^\gamma}.$$  

After the FSR event is generated with crude distribution, it is only accepted
if the photon polar angle is in the interval 
$-c_m\le\cos{\theta}\le c_m$. Otherwise the procedure is repeated.

We also need the total crude FSR cross section. It can be obtained by 
integrating (\ref{fsrdw}), but there is some subtlety in how to 
account for the photon angular cut. Note that
$$\int \frac{1}{y_1y_2}\delta (p_-+p_+-k-q_--q_+)~\frac{d\vec{q}_-}{E_-}
~\frac{d\vec{q}_+}{E_+}$$
can depend only on $k\cdot (p_-+p_+)$ and not separately on $k\cdot p_-$
and/or $k\cdot p_+$. Therefore in the CMS frame $d\sigma_0^{(FSR)}/d\vec{k}$
does not depend on the photon angular variables. Hence the distribution over
$\cos{\theta}$ is uniform and the desired total cross section is given by:
\begin{equation}
\sigma_0^{(FSR)}=\frac{2\alpha^3 c_m}{s}\int\limits_{x_{min}}^{1-\mu^2}
\ln{\frac{2(1-x)-\mu^2+2\sqrt{(1-x)(1-x-\mu^2)}}{\mu^2}}~\frac{dx}{x}.
\label{sg0FSR} \end{equation}
This integral is calculated numerically in the MUMUG program.

\section{Concluding remarks}
In the present paper, we have described and made available MUMUG, a new
leading-order Monte Carlo program that simulates  $e^+ e^- \to \mu^+ 
\mu^- \gamma$ process.

We hope that the MUMUG program in conjunction  with this article can be used
for educational purposes. Therefore, we have presented a very detailed 
description of the Monte Carlo algorithm used, as well as the helicity 
amplitude method and the calculation of the $e^+ e^- \to \mu^+ \mu^- \gamma$ 
matrix element by means of this method.

For reference purposes, in Table \ref{Tab1} we provide some total cross 
sections calculated by using the MUMUG program. For comparison with other 
programs, keep in mind that professional Monte Carlo programs such as AFKQED 
or MCGPJ include the emission of additional collinear photons, while others, 
such as PHOKHARA and KKMC, include next-to-leading order radiative corrections.
All these improvements are missing in the MUMUG generator. 
\begin{table}[htb]
\begin{tabular}{|c|c|c|c|c|}
\hline
\backslashbox{$\omega_{min}$}{$\theta_{min}$} & $10^\circ$ & $15^\circ$ & $20^\circ$ 
& $25^\circ$ \\ \hline 0.1~GeV & 206.20$\pm$ 0.18 & 182.83$\pm$ 0.16 & 164.60
$\pm$ 0.14 & 149.24$\pm$ 0.13 \\ \hline 0.15~GeV & 187.54 $\pm$ 0.16 &
165.62 $\pm$ 0.14 & 148.70 $\pm$ 0.13 & 134.43 $\pm$ 0.12 \\ \hline
0.20~GeV & 174.35 $\pm$ 0.15 & 153.43 $\pm$ 0.13 & 137.56 $\pm$ 0.12 &
124.08 $\pm$ 0.11 \\ \hline 0.25~GeV & 164.28 $\pm$ 0.14 & 143.99 $\pm$ 0.12
& 128.86 $\pm$ 0.11 & 116.14 $\pm$ 0.10 \\ \hline 
\end{tabular}
\caption{The total cross sections of $e^+ e^- \to \mu^+ \mu^- \gamma$ in
picobarns for the CMS beam energy $E=5.29~\mathrm{GeV}$ and for different 
values of the minimum photon energy ($\omega_{min}$) and angular cuts imposed
on the photon polar angle ($\theta_{min}$).}
\label{Tab1}
\end{table}

The  Fortran code of the MUMUG program can be downloaded (under the
MIT license) from: \url{https://wwwsnd.inp.nsk.su/~silagadz/MUMUG/}, or from 
the Code Ocean Capsule:  \url{https://codeocean.com/capsule/4243841/tree/v1}.

\begin{acknowledgements}
The work is supported by the Ministry of Education and Science of the Russian 
Federation and in part by RFBR grant 20-02-00697-a.
\end{acknowledgements}
 
\appendix
\section*{Appendix}
\section{Feynman rules used in the main text}
Each Feynman diagram determines a contribution to the momentum space
matrix element of the process under consideration. This contribution
can be found from the diagram according to the following rules (only
ones relevant for our studies are listed). Their derivations can be 
found in the standard quantum field theory textbooks \cite{QFT}. 

\begin{itemize}
\item Initial state particles with 4-momentum $p$:  

\begin{picture}(20000,6000)
\put(4000,4500){electron}
\drawline\fermion[\E\REG](10000,4500)[6000]
\global\advance\pmidy by -350
\put(\pmidx,\pmidy){$>$}
\put(\pbackx,\pbacky){\circle*{500}}
\global\advance\pbackx by 2000
\put(\pbackx,\pbacky){$u_\lambda (p)$}
\global\advance\pbacky by -3000
\put(4000,\pbacky){positron}
\drawline\fermion[\E\REG](10000,\pbacky)[6000]
\global\advance\pmidy by -350
\put(\pmidx,\pmidy){$<$}
\put(\pbackx,\pbacky){\circle*{500}}
\global\advance\pbackx by 2000
\put(\pbackx,\pbacky){$\bar v_\lambda (p)$}
\end{picture}

\noindent where $\lambda$ stands for the fermion polarization. 
$u_\lambda(p)$ and $v_\lambda(p)$ are particle and antiparticle Dirac spinors.

\item Final state particles with 4-momentum $p$:

\begin{picture}(20000,10000)
\put(4000,8000){electron}
\drawline\fermion[\E\REG](10000,8000)[6000]
\global\advance\pmidy by -350
\put(\pmidx,\pmidy){$>$}
\put(\pfrontx,\pfronty){\circle*{500}}
\global\advance\pbackx by 2000
\put(\pbackx,\pbacky){$\bar u_\lambda (p)$}
\global\advance\pbacky by -3000
\put(4000,\pbacky){positron}
\drawline\fermion[\E\REG](10000,\pbacky)[6000]
\global\advance\pmidy by -350
\put(\pmidx,\pmidy){$<$}
\put(\pfrontx,\pfronty){\circle*{500}}
\global\advance\pbackx by 2000
\put(\pbackx,\pbacky){$ v_\lambda (p)$}
\global\advance\pbacky by -3000
\put(4000,\pbacky){photon}
\drawline\photon[\E\REG](10000,\pbacky)[6]
\global\advance\pmidy by -350
\global\advance\pmidx by 250
\put(\pmidx,\pmidy){$\big >$}
\put(\pfrontx,\pfronty){\circle*{500}}
\global\advance\pbackx by 2000
\put(\pbackx,\pbacky){$\epsilon^*_\mu(p)$}
\end{picture}

$\epsilon_\mu$ is the photon polarization four-vector.

\item Intermediate state particle propagators with 4-momentum $p$:

\begin{picture}(20000,7000)
\put(4000,5000){fermion}
\drawline\fermion[\E\REG](10000,5500)[6000]
\put(\pfrontx,\pfronty){\circle*{500}}
\put(\pbackx,\pbacky){\circle*{500}}
\global\advance\pmidy by -350
\put(\pmidx,\pmidy){$>$}
\global\advance\pbackx by 4000
\put(\pbackx,\pbacky){$\frac{i(\hat p +m)}{p^2-m^2+i\epsilon}$}
\global\advance\pbacky by -4000
\put(4000,\pbacky){photon}
\drawline\photon[\E\REG](10000,\pbacky)[6]
\put(\pfrontx,\pfronty){\circle*{500}}
\put(\pbackx,\pbacky){\circle*{500}}
\global\advance\pmidy by -350
\global\advance\pmidx by 250
\put(\pmidx,\pmidy){$\big >$}
\global\advance\pbackx by 5000
\put(\pbackx,\pbacky){$\frac{-ig_{\mu\nu}}{p^2+i\epsilon}$}
\end{picture}

\item Interaction vertex:

\begin{picture}(20000,6500)
\drawline\fermion[\SE\REG](10000,5000)[4000]
\drawarrow[\NW\ATBASE](\pmidx,\pmidy)
\put(\pbackx,\pbacky){\circle*{500}}
\drawline\photon[\E\REG](\pbackx,\pbacky)[6]
\drawline\fermion[\SW\REG](\pfrontx,\pfronty)[4000]
\drawarrow[\NE\ATBASE](\pmidx,\pmidy)
\global\advance\photonbackx by 3000
\put(\photonbackx,\photonbacky){$-ie\gamma_\mu$}
\end{picture}

\end{itemize}

\section{Usage of the program}
To generate pseudorandom numbers uniformly distributed  in the interval 
$(0,1)$, the program uses $RANLUX$ generator from the CERNLIB library
\cite{RANLUX}. Therefore, if the user does not want to use default values
for $RANLUX$, $RLUXGO$ subroutine should be called to initialize $RANLUX$,
as described in \cite{RANLUX}.

Various internal common blocks of MUMUG are filled when 
$$FILCOM(E,EGMIN,THGMIN)$$ subroutine is called. Here $E (REAL*8)$ is the CMS 
beam energy in GeV. $EGMIN (REAL*8)$ represents the minimum energy of the 
photon, also in GeV, and $THGMIN (REAL*8)$ is the angular cut on the photon 
polar angle in degrees.
    
After $RANLUX$ generator was initialized, and $FILCOM$ subroutine was called,
MUMUG is ready to generate events of the process $e^+ e^- \to \mu^+ 
\mu^- \gamma$. Each call of the $$MUMUG(E,QM,QP,K,IMOD)$$ subroutine results 
in the output of $REAL*8$ arrays $QM(0:3),QP(0:3),K(0:3)$, which represent 
four-momenta of the produced particles ($\mu^+$, $\mu^-$ and $\gamma$, 
respectively).
 
The $IMOD (INTEGER)$ parameter specifies which expression is used by MUMUG
when calculating the matrix element of $e^+ e^- \to \mu^+ \mu^- \gamma$  (or 
its modulus squared). In particular, if $IMOD=0$, expression $(12)$ (with the 
corrected sign misprint) from \cite{AFKM} is used.  If $IMOD=1$, expressions 
$(9--10)$ from \cite{AFKM} are used. If $IMOD=2$, the results of \cite{BKJW} 
are used. At last, if $IMOD=3$, our helicity amplitudes, described in this 
article, are used. These different modes of matrix element calculation have 
been implemented in MUMUG in order to facilitate the detection of various 
programming mistakes (bugs) during the program development. All modes produce 
identical output (within the precision expected by the double precision 
arithmetic used).
 
After sufficiently large number of $MUMUG$ calls, $crs(E,sg,sgerr)$ 
subroutine, if called, estimates the cross section $sg (REAL*8)$ of the process
$e^+ e^- \to \mu^+ \mu^- \gamma$, $E_\gamma>EGMIN$, $THGMIN<\theta_\gamma<
180^\circ-THGMIN$, and its error $sgerr (REAL*8)$ in picobarns.

\end{document}

%% file: FEYNMAN.tex
%
%
%
%
%
%
\message{FEYNMAN:  For generating Feynman Diagrams in LaTex}
\message{Mark 1.0 Last Altered by MJSL 2/89}
\textheight 650pt \textwidth 400pt  
\setlength{\unitlength}{0.01pt}
\gdef\Feynmanlength{\setlength{\unitlength}{0.01pt}}  
\gdef\unlock{\catcode`\@=11}
\gdef\lock{\catcode`\@=12}
\global\newcount\LINETYPE                     
\global\newcount\LINEDIRECTION
\global\newcount\LINECONFIGURATION
\newcommand{\LTYPE}{\LINETYPE}
\newcommand{\LDIR}{\LINEDIRECTION}
\newcommand{\LCONFIG}{\LINECONFIGURATION}
\global\LINETYPE=1  \global\LINEDIRECTION=0  \global\LINECONFIGURATION=0
\global\newcount\fermion    \fermion=1
\global\newcount\scalar     \scalar=2
\global\newcount\photon     \photon=3
\global\newcount\gluon      \gluon=4
\global\newcount\especial   \especial=5
\gdef\N{0}  \gdef\NE{1}  \gdef\E{2}   \gdef\SE{3}
\gdef\S{4}  \gdef\SW{5}  \gdef\W{6}   \gdef\NW{7}
\global\newcount\REG            \global\REG=0
\global\newcount\FLIPPED        \global\FLIPPED=1
\global\newcount\CURLY          \global\CURLY=2
\global\newcount\FLIPPEDCURLY   \global\FLIPPEDCURLY=3
\global\newcount\FLAT           \global\FLAT=4
\global\newcount\FLIPPEDFLAT    \global\FLIPPEDFLAT=5
\global\newcount\CENTRAL        \global\CENTRAL=6
\global\newcount\FLIPPEDCENTRAL \global\FLIPPEDCENTRAL=7
\gdef\LONGPHOTON{6}             \gdef\FLIPPEDLONG{7}
\global\newcount\SQUASHEDGLUON  \global\SQUASHEDGLUON=8
\gdef\SQUASHED{\SQUASHEDGLUON}
%
\newcount\adjx \adjx=0
\newcount\adjy \adjy=0
\global\newdimen\BIGPHOTONS     \BIGPHOTONS=0pt  
\gdef\bigphotons{\global\BIGPHOTONS=12pt}
\global\newdimen\THICKPHOTONS     \THICKPHOTONS=0pt  
\global\newdimen\THICKPHOTONSWITCH    \THICKPHOTONSWITCH=0pt
\gdef\THICKPHOTONTEST{
\THICKPHOTONSWITCH=0pt
\ifdim\THICKPHOTONS=0pt \relax
  \else \ifnum\LTYPE=3
           \ifnum\LDIR=2 \THICKPHOTONSWITCH=1pt \fi 
           \ifnum\LDIR=6 \THICKPHOTONSWITCH=1pt \fi 
        \fi
\fi
}  
\gdef\THICKLINES{\thicklines  \THICKPHOTONS=1pt}
\gdef\THINLINES{\thinlines  \THICKPHOTONS=0pt}
\global\newcount\phantomswitch   \global\phantomswitch=0
\global\newcount\stemlength   \global\stemlength=275   
\global\newcount\absstemlength        
\global\newcount\stemlengthx          
\global\newcount\stemlengthy          
\newdimen\FRONTSTEM  \FRONTSTEM=0pt   
\newdimen\BACKSTEM   \BACKSTEM=0pt    
\newdimen\EITHERSTEM \EITHERSTEM=0pt  
\gdef\frontstemmed{\FRONTSTEM=1pt}            
\gdef\backstemmed{\BACKSTEM=1pt}              
\gdef\stemmed{\FRONTSTEM=1pt  \BACKSTEM=1pt}    
\global\newcount\arrowlength                
\global\newdimen\ATTIP   \global\ATTIP=0pt  
\global\newdimen\ATBASE  \global\ATBASE=1pt 
\global\newcount\unitboxnumber  
\global\newcount\unitboxnumberpo  
\global\newcount\particlelengthx  
\gdef\plengthx{\particlelengthx}
\global\newcount\particlelengthy  
\gdef\plengthy{\particlelengthy}  
\global\newcount\boxlengthx  
\global\newcount\boxlengthy  
\global\newcount\particleadjustx  
\global\newcount\particleadjusty  
\global\newcount\particlelength   
\global\newcount\particlefrontx
\gdef\pfrontx{\particlefrontx}
\global\newcount\PFRONTx
\global\newcount\particlefronty
\gdef\pfronty{\particlefronty}
\global\newcount\PFRONTy
\global\newcount\particlebackx
\gdef\pbackx{\particlebackx}
\global\newcount\particlebacky
\gdef\pbacky{\particlebacky}
\global\newcount\particlemidx
\gdef\pmidx{\particlemidx}
\global\newcount\particlemidy
\gdef\pmidy{\particlemidy}
\global\newcount\seglength  \global\newcount\gaplength
\global\gaplength=850  
\global\seglength=1416  
\global\newcount\Xone    \global\newcount\Yone    
\global\newcount\Xtwo    \global\newcount\Ytwo    
\global\newcount\Xthree  \global\newcount\Ythree  
\global\newcount\Xfour   \global\newcount\Yfour   
\global\newcount\Xfive   \global\newcount\Yfive   
\global\newcount\Xsix    \global\newcount\Ysix    
\global\newcount\Xseven  \global\newcount\Yseven  
\global\newcount\Xeight  \global\newcount\Yeight  
%
%
\newsavebox{\lastline}  
\global\newcount\numlineparts   
\global\newcount\upperlineadjx  \upperlineadjx=0  
\global\newcount\upperlineadjy  \upperlineadjy=0  
\global\newcount\lowerlineadjx  \lowerlineadjx=0  
\global\newcount\lowerlineadjy  \lowerlineadjy=0  
\global\newcount\thirdlineadjx  \thirdlineadjx=0  
\global\newcount\thirdlineadjy  \thirdlineadjy=0  
\global\newcount\fourthlineadjx \fourthlineadjx=0  
\global\newcount\fourthlineadjy \fourthlineadjy=0  
\global\newcount\unitboxwidth   \unitboxwidth=1000
\global\newcount\unitboxheight  \unitboxheight=0  
\global\newcount\numupperunits  \numupperunits=8  
\global\newcount\numlowerunits  \numlowerunits=8  
\global\newcount\numthirdunits  \numthirdunits=8  
\global\newcount\numfourthunits \numfourthunits=8  
\global\newcount\fermioncount   \global\fermioncount=0    
\global\newcount\scalarcount    \global\scalarcount=0    
\global\newcount\photoncount    \global\photoncount=0    
\global\newcount\gluoncount     \global\gluoncount=0    
\global\newcount\especialcount  \global\especialcount=0    
\global\newcount\vertexcount    \global\vertexcount=-1
%
\global\newcount\XDIR
\global\newcount\YDIR
\gdef\SETDIR{  
\ifcase\LDIR 
     \global\XDIR=0  \global\YDIR=1   
\or  \global\XDIR=1  \global\YDIR=1   
\or  \global\XDIR=1  \global\YDIR=0   
\or  \global\XDIR=1  \global\YDIR=-1  
\or  \global\XDIR=0  \global\YDIR=-1  
\or  \global\XDIR=-1 \global\YDIR=-1  
\or  \global\XDIR=-1 \global\YDIR=0   
\or  \global\XDIR=-1 \global\YDIR=1   
\else\DIRECTERROR 
\fi}  
\gdef\moduloeight#1{
\ifnum#1>7 \global\advance #1 by -8 
\relax
\moduloeight#1 
\relax
\else \relax  
\fi}
\gdef\multroothalf#1{\global\multiply #1 by 7071 \global\divide #1 by 10000}
\gdef\negate#1{\global\multiply #1 by -1}
\gdef\double#1{\global\multiply #1 by 2}
\gdef\slanttest(#1,#2){ 
\ifodd\LDIR
\multiply #1 by 7071  \divide #1 by 10000
\multiply #2 by 7071  \divide #2 by 10000
\fi
}
\gdef\gslanttest(#1,#2){
\ifodd\LDIR
\multroothalf#1
\multroothalf#2
\fi
}
%
%
\gdef\setplength{ 
\global\particlelengthx=\unitboxwidth
\global\particlelengthy=\unitboxheight
\global\multiply \particlelengthx by \unitboxnumber
\global\multiply \particlelengthy by \unitboxnumber
\global\advance \particlelengthx by \particleadjustx
\global\advance \particlelengthy by \particleadjusty
}
\gdef\boxlengthdefault{  
\global\boxlengthx=\plengthx
\global\boxlengthy=\plengthy
\ifnum\plengthx<0 \global\multiply\boxlengthx by -1 \fi
\ifnum\plengthy<0 \global\multiply\boxlengthy by -1 \fi
}
\gdef\rearcoords{  
\global\particlebacky=\particlefronty 
\global\particlebackx=\particlefrontx 
\global\advance \particlebackx by \particlelengthx
\global\advance \particlebacky by \particlelengthy
}
\gdef\midcoords{  
\global\particlemidy=\particlefronty
\global\particlemidx=\particlefrontx
\global\stemlengthx=\particlelengthx  
\global\stemlengthy=\particlelengthy  
\global\divide\stemlengthx by 2
\global\divide\stemlengthy by 2
\global\advance \particlemidx by \stemlengthx
\global\advance \particlemidy by \stemlengthy
}
\gdef\setparticle{\setplength\rearcoords\midcoords\boxlengthdefault}  
\gdef\setcoords(#1,#2,#3)(#4,#5,#6)[#7,#8]{  
\global\upperlineadjx=#1
\global\lowerlineadjx=#2
\global\thirdlineadjx=#3
\global\upperlineadjy=#4
\global\lowerlineadjy=#5
\global\thirdlineadjy=#6
\global\unitboxwidth=#7
\global\unitboxheight=#8
}
%
%
%
\gdef\drawoldpic#1(#2,#3){  
\global\particlefrontx=#2
\global\particlefronty=#3
\rearcoords  
\midcoords
\put(#2,#3){\usebox{#1}}
}
\gdef\drawsavedline`#1' as #2[#3#4](#5,#6)[#7]{
\global\LINETYPE=#2
\global\LINEDIRECTION=#3
\global\LINECONFIGURATION=#4
\global\particlefrontx=#5
\global\particlefronty=#6
\global\unitboxnumber=#7  
\selectcase
\rearcoords
\midcoords
\ifnum\phantomswitch=0 \drawas{#1}\fi
}

\gdef\startphantom{\phantomswitch=1} 
\gdef\stopphantom{\phantomswitch=0}  

\gdef\drawas#1{
\global\savebox{#1}(\boxlengthx,\boxlengthy){
\setlength{\unitlength}{0.01pt}
\begin{picture}(\boxlengthx,\boxlengthy)
\multiput(\upperlineadjx,\upperlineadjy)(\unitboxwidth,\unitboxheight)
{\numupperunits}{\upperunitbox}
\ifnum\numlineparts > 1  
\multiput(\lowerlineadjx,\lowerlineadjy)(\unitboxwidth,\unitboxheight)
{\numlowerunits}{\lowerunitbox}  
\fi
\ifnum\numlineparts > 2  
\multiput(\thirdlineadjx,\thirdlineadjy)(\unitboxwidth,\unitboxheight)
{\numthirdunits}{\thirdunitbox}  
\fi
\ifnum\numlineparts > 3  
\multiput(\fourthlineadjx,\fourthlineadjy)(\unitboxwidth,\unitboxheight)
{\numfourthunits}{\lowerunitbox}  
\fi
\end{picture} }
\global\PFRONTx=\pfrontx  \global\PFRONTy=\pfronty   
\SETFRONTSTEM
\THICKPHOTONTEST
\ifdim\THICKPHOTONSWITCH=1pt\global\advance\PFRONTy by 20  \fi
\put(\PFRONTx,\PFRONTy) {\usebox{#1}}   
\ifdim\THICKPHOTONSWITCH=1pt
\global\advance\PFRONTy by -40
\put(\PFRONTx,\PFRONTy) {\usebox{#1}}   
\global\advance \PFRONTy by 20  
\fi  
\SETBACKSTEM
\seglength=1416   \gaplength=850   
}
%
%

\gdef\drawandsaveline`#1' as #2[#3#4](#5,#6)[#7]{
\global\newsavebox{#1}
\drawsavedline`#1' as #2[#3#4](#5,#6)[#7]
}

\gdef\drawline#1[#2#3](#4,#5)[#6]{   
\drawsavedline`\lastline' as #1[#2#3](#4,#5)[#6]}

\gdef\saveas#1{  
\global\newsavebox#1
\drawas#1}
%
%
%
\gdef\TYPEERROR{\message{*** ERROR IN PARTICLE TYPE SELECTION ***}
\message{+++ Try with line type \fermion,\scalar,\photon,\gluon 
(see manual) +++}\SETERR}
\gdef\DIRECTERROR{\SETERR\message{*** ERROR IN PARTICLE DIRECTION SELECTION ***}
\message{+++ Try again with direction N, NE, E, SE  etc. or see manual +++}}
\gdef\UNIMPERROR{\message{*** ERROR IN PARTICLE OPTIONS SELECTION ***}
\message{
+++ The requested options combination has not yet been implemented +++}\SETERR}
\gdef\SETERR{\gdef\upperunitbox{{\tiny Error}}  
\gdef\lowerunitbox{\relax}
\gdef\thirdunitbox{\relax}
}
\gdef\neglengthcheck{\ifnum\unitboxnumber < 1 
\message{   *** ERROR:  PARTICLE OF NEGATIVE OR ZERO LENGTH REQUESTED. ***   }
\message{   ***         TAKING ABSOLUTE VALUE. ***   }\negate\unitboxnumber \fi}
\gdef\selectcase{  
\neglengthcheck   
\SETDIR  
\ifcase\LINETYPE
\TYPEERROR  
\or \selectfermion  
\or \selectscalar   
\or \selectphoton   
\or \selectgluon    
\or \selectespecial 
\else \TYPEERROR \fi  }
\gdef\selectfermion{
\ifnum\fermioncount=0 \input fermionsetup \fi   
\global\advance\fermioncount by 1  
\ALLfermion   
}
\gdef\selectscalar{
\ifnum\scalarcount=0 \input scalarsetup \fi   
\global\advance\scalarcount by 1  
\ALLscalar
}
\gdef\selectphoton{   
\ifnum\photoncount=0 \input photonsetup  \fi   
\selectphoton
}
\gdef\selectgluon{   
\ifnum\gluoncount=0 \input gluonsetup  \fi
\selectgluon
}
\gdef\selectespecial{\UNIMPERROR}
%
%
\gdef\checkvertex{ 
\ifnum\vertexcount=-1   \input vertex  \fi}
\gdef\drawvertex#1[#2#3](#4,#5)[#6]{\checkvertex\drawvertex#1[#2#3](#4,#5)[#6]}
\gdef\vertexcap#1{\checkvertex\vertexcap#1}
\gdef\vertexcaps{\checkvertex\vertexcaps}
\gdef\vertexlink#1{\checkvertex\vertexlink#1}
\gdef\vertexlinks{\checkvertex\vertexlinks}
\gdef\stemvertex#1{\checkvertex\stemvertex#1}
\gdef\stemvertices{\checkvertex\stemvertices}
\gdef\flipvertex{\checkvertex\flipvertex}
%
%
\global\arrowlength=349  
\gdef\drawarrow[#1#2](#3,#4){
\global\LDIR=#1
\SETDIR
\global\boxlengthx=#3  
\global\boxlengthy=#4  
\ifdim#2=1pt  
\adjx=\arrowlength      \adjy=\arrowlength
\multiply\adjx by \XDIR \multiply\adjy by \YDIR  
\slanttest(\adjx,\adjy)
\global\advance\boxlengthx by \adjx    \global\advance\boxlengthy by \adjy
\fi
\ifnum\phantomswitch=0\put(\boxlengthx,\boxlengthy){\vector(\XDIR,\YDIR){0}}\fi
}  
%
%
\gdef\SETFRONTSTEM{
\EITHERSTEM=\FRONTSTEM   \advance\EITHERSTEM by \BACKSTEM
\ifdim\EITHERSTEM>0pt
\global\stemlengthx=\stemlength   \global\stemlengthy=\stemlength   
\global\absstemlength=\stemlength   
\SETDIR
\gslanttest(\stemlengthx,\stemlengthy)
\gslanttest(\absstemlength,\REG)  
\ifnum\XDIR=0 \stemlengthx=0 \fi
\ifnum\YDIR=0 \stemlengthy=0 \fi
\global\multiply\stemlengthx by \XDIR
\global\multiply\stemlengthy by \YDIR
\ifdim\FRONTSTEM=1pt 
\ifnum\phantomswitch=0
          \put(\pfrontx,\pfronty){\line(\XDIR,\YDIR){\absstemlength}}\fi
\global\advance\plengthx by \stemlengthx
\global\advance\plengthy by \stemlengthy
\global\advance\PFRONTx by \stemlengthx   
\global\advance\PFRONTy by \stemlengthy
\global\advance\pmidx by \stemlengthx
\global\advance\pmidy by \stemlengthy
\global\advance\pbackx by \stemlengthx
\global\advance\pbacky by \stemlengthy
\ifnum\LTYPE=3
\global\photonfrontx=\PFRONTx  \global\photonfronty=\PFRONTy
\global\photonbackx=\pbackx    \global\photonbacky=\pbacky
\fi  
\ifnum\LTYPE=4
\global\gluonfrontx=\PFRONTx  \global\gluonfronty=\PFRONTy
\global\gluonbackx=\pbackx    \global\gluonbacky=\pbacky
\fi  
\fi  
\fi  
}    
\gdef\SETBACKSTEM{
\ifdim\BACKSTEM=1pt 
\ifnum\phantomswitch=0
       \put(\pbackx,\pbacky){\line(\XDIR,\YDIR){\absstemlength}}\fi
\global\advance\plengthx by \stemlengthx
\global\advance\plengthy by \stemlengthy
\global\advance\pbackx by \stemlengthx
\global\advance\pbacky by \stemlengthy
\fi  
\global\stemlength=275  \FRONTSTEM=0pt  \BACKSTEM=0pt 
}    
\gdef\drawloop#1[#2#3](#4,#5){  
\input loops  
\drawloop#1[#2#3](#4,#5)}
\Feynmanlength  

%% file: fermionsetup.tex
\global\newcount\fermionlength  
\global\newcount\fermionlengthx
\global\newcount\fermionlengthy
\global\newcount\fermionfrontx  
\global\newcount\fermionfronty  
\global\newcount\fermionbackx
\global\newcount\fermionbacky
\gdef\ALLfermion{  
\global\fermionfrontx=\particlefrontx \global\fermionfronty=\particlefronty
\ifnum\unitboxnumber > 50000
\message{   *** WARNING *** Fermion of length
\the\unitboxnumber\space requested ***   }
\ifnum\unitboxnumber > 80000
\message{   *** Reducing fermion length to 30000 (max 80000) ***   }
\global\unitboxnumber=30000 \fi \fi  
\global\fermionlength=\unitboxnumber 
\global\particleadjustx=0   \global\particleadjusty=0 
\global\numlineparts = 1    \global\numupperunits=1
\global\upperlineadjx=-200  \global\upperlineadjy=0
\global\fermionlengthx=\fermionlength    \global\fermionlengthy=\fermionlength
\gslanttest(\fermionlengthx,\fermionlengthy)  
\global\multiply\fermionlengthx by \XDIR  
\global\multiply\fermionlengthy by \YDIR  
\global\unitboxheight=\fermionlengthy   \global\unitboxwidth=\fermionlengthx   
\global\advance \fermionlengthx by \particleadjustx
\global\advance \fermionlengthy by \particleadjusty
\global\particlelengthx=\fermionlengthx
\global\particlelengthy=\fermionlengthy  
\boxlengthdefault    \rearcoords    \midcoords
\global\fermionbackx=\particlebackx     \global\fermionbacky=\particlebacky
\ifcase\LINECONFIGURATION  
\ifnum\XDIR=0 
\gdef\upperunitbox{\line(\XDIR,\YDIR){\boxlengthy}} 
\else
\gdef\upperunitbox{\line(\XDIR,\YDIR){\boxlengthx}}
\fi
\else \UNIMPERROR
\fi
}

%% file: photonsetup.tex
\newcount\numwiggles    \newcount\numwigglespo
\global\newcount\photonlengthx
\global\newcount\photonlengthy
\global\newcount\photonfrontx  
\global\newcount\photonfronty  
\global\newcount\photonbackx
\global\newcount\photonbacky
\newcount\halfwigglelength
\global\font\Twelverom=cmr12
\global\font\Tenrom=cmr10
\gdef\Lbr{{\Twelverom(}}   \gdef\Rbr{{\Twelverom)}}
\gdef\SLbr{{\Tenrom(}}     \gdef\SRbr{{\Tenrom)}}
\gdef\Smile{{\large$\smile$}}  
\gdef\Frown{{\large$\frown$}}  
\ifdim\BIGPHOTONS>0pt  \gdef\Smile{$\smile$} \gdef\Frown{$\frown$} \fi
%
\gdef\selectphoton{   
\global\advance\photoncount by 1  
\global\photonfrontx=\particlefrontx   
\global\photonfronty=\particlefronty   
\ifnum\unitboxnumber > 50
\message{   *** WARNING *** Photon with 
\the\unitboxnumber\space half-wiggles requested ***   }
\ifnum\unitboxnumber > 150
\message{   *** Reducing photon length to 10 half-wiggles (max 150) ***   }
\ifnum\unitboxnumber > 1000
\message{   *** Probable Cause:  Photon selected instead of Fermion ***   }
\fi \global\unitboxnumber=10 \fi \fi  
\numwiggles=\unitboxnumber
\divide\numwiggles by 2
\global\unitboxnumberpo=\numwiggles 
\global\multiply \unitboxnumberpo by -1
\numwigglespo=\unitboxnumber
\advance\numwigglespo by \unitboxnumberpo 
\global\numlineparts = 2  
\global\numupperunits=\numwigglespo  
\global\numlowerunits=\numwiggles  
\particleadjustx=0  
\particleadjusty=0  
\ifcase\LINEDIRECTION
     \Nphoton    
\or  \NEphoton   
\or  \Ephoton    
\or  \SEphoton   
\or  \Sphoton    
\or  \SWphoton   
\or  \Wphoton    
\or  \NWphoton   
\else\DIRECTERROR \fi
\setplength
\global\divide\plengthx by 2  \global\divide\plengthy by 2
\rearcoords  \boxlengthdefault   \midcoords
\global\photonbackx=\pbackx  
\global\photonbacky=\pbacky  
\global\photonlengthx=\plengthx  
\global\photonlengthy=\plengthy  
}
\gdef\SETUNITBOX(#1)[#2][#3]{ 
\gdef\upperunitbox{\oval(#1,#1)[#2]}
\gdef\lowerunitbox{\oval(#1,#1)[#3]}
}
\gdef\Nphoton{  
\ifcase\LINECONFIGURATION  
\setcoords(-490,-250,0)(260,1250,0)[0,2000]
\gdef\upperunitbox{\SLbr}   \gdef\lowerunitbox{\SRbr}
\particleadjusty=10
\or 
\setcoords(-271,-501,0)(250,1250,0)[0,2000]   
\gdef\upperunitbox{\SRbr}   \gdef\lowerunitbox{\SLbr}
\or 
\particleadjusty=0
\setcoords(-501,-351,0)(300,1400,0)[0,2200]
\gdef\upperunitbox{\Lbr}   \gdef\lowerunitbox{\Rbr}
\or 
\setcoords(-353,-499,0)(300,1400,0)[0,2200]
\gdef\upperunitbox{\Rbr}   \gdef\lowerunitbox{\Lbr}
\or 
\setcoords(-481,-371,0)(280,1300,0)[0,2000]
\gdef\upperunitbox{\Lbr}   \gdef\lowerunitbox{\Rbr}
\particleadjusty=150
\ifnum\numwiggles=\number\numwigglespo \particleadjustx=-50 \fi
\or 
\setcoords(-321,-391,0)(280,1300,0)[0,2000]
\gdef\upperunitbox{\Rbr}   \gdef\lowerunitbox{\Lbr}
\particleadjusty=150
\ifnum\numwiggles=\number\numwigglespo \particleadjustx=80 \fi
\or 
\setcoords(-490,-260,0)(300,1500,0)[0,2400]
\gdef\upperunitbox{\Lbr}   \gdef\lowerunitbox{\Rbr}
\or 
\setcoords(-301,-531,0)(300,1500,0)[0,2400]
\gdef\upperunitbox{\Rbr}   \gdef\lowerunitbox{\Lbr}
\else \UNIMPERROR
\fi
}
\gdef\NEphoton{    
\ifcase\LINECONFIGURATION  
\setcoords(425,425,0)(1250,0,0)[1250,1250]       \SETUNITBOX(1250)[br][tl]  
\ifnum\numwigglespo > \number \numwiggles \particleadjustx=15 \fi
\or 
\setcoords(1050,-200,0)(625,625,0)[1250,1250]    \SETUNITBOX(1250)[tl][br]
\ifnum\numwigglespo > \number \numwiggles \particleadjustx=25 \fi
\or 
\setcoords(500,500,0)(1400,0,0)[1400,1400]       \SETUNITBOX(1400)[br][tl]
\or 
\setcoords(1200,-200,0)(700,700,0)[1400,1400]    \SETUNITBOX(1400)[tl][br]  
\or 
\setcoords(400,400,0)(1200,0,0)[1200,1200]       \SETUNITBOX(1200)[br][tl]  
\or 
\setcoords(1000,-200,0)(600,600,0)[1200,1200]    \SETUNITBOX(1200)[tl][br]
\else \UNIMPERROR
\fi
\numupperunits=\numwiggles   \numlowerunits=\numwigglespo
}
\gdef\Ephoton{    
\ifcase\LINECONFIGURATION  
\setcoords(-285,715,0)(-150,-400,0)[2005,0]
\gdef\upperunitbox{\Frown}   \gdef\lowerunitbox{\Smile}
\or  
\setcoords(-285,715,0)(-420,-170,0)[2005,0]
\gdef\upperunitbox{\Smile}   \gdef\lowerunitbox{\Frown}
\else \UNIMPERROR
\fi
\particleadjustx=-15 
}
\gdef\SEphoton{   
\ifcase\LINECONFIGURATION  
\setcoords(-200,1050,0)(-625,-625,0)[1250,-1250] \SETUNITBOX(1250)[tr][bl]
\ifnum\numwigglespo > \number \numwiggles \particleadjustx=25 \fi
\or 
\setcoords(425,425,0)(0,-1250,0)[1250,-1250]     \SETUNITBOX(1250)[bl][tr]
\ifnum\numwigglespo > \number \numwiggles \particleadjustx=15 \fi
\or 
\setcoords(-200,1200,0)(-700,-700,0)[1400,-1400] \SETUNITBOX(1400)[tr][bl]  
\or 
\setcoords(500,500,0)(0,-1400,0)[1400,-1400]     \SETUNITBOX(1400)[bl][tr]  
\or 
\setcoords(-200,1000,0)(-600,-600,0)[1200,-1200] \SETUNITBOX(1200)[tr][bl]
\particleadjustx=-20
\or 
\setcoords(420,420,0)(0,-1200,0)[1200,-1200]     \SETUNITBOX(1200)[bl][tr]
\particleadjustx=40
\else \UNIMPERROR
\fi
}
\gdef\Sphoton{  
\ifcase\LINECONFIGURATION  
\setcoords(-252,-490,0)(-740,-1740,0)[0,-2000]
\gdef\upperunitbox{\SRbr}   \gdef\lowerunitbox{\SLbr}
\or 
\setcoords(-490,-260,0)(-740,-1740,0)[0,-2002]
\gdef\upperunitbox{\SLbr}   \gdef\lowerunitbox{\SRbr}
\or 
\setcoords(-299,-449,0)(-870,-1970,0)[0,-2200]
\gdef\upperunitbox{\Rbr}    \gdef\lowerunitbox{\Lbr}
\particleadjusty=-95
\or 
\setcoords(-517,-371,0)(-900,-2000,0)[0,-2200]
\gdef\upperunitbox{\Lbr}    \gdef\lowerunitbox{\Rbr}
\particleadjusty=-165
\or 
\setcoords(-299,-409,0)(-885,-1905,0)[0,-2000]
\gdef\upperunitbox{\Rbr}   \gdef\lowerunitbox{\Lbr}
\particleadjustx=50     \particleadjusty=-380
\ifodd\unitboxnumber\relax\else\particleadjustx=250 \particleadjusty=-400 \fi
\or 
\setcoords(-519,-449,0)(-900,-1920,0)[0,-2000]
\gdef\upperunitbox{\Lbr}   \gdef\lowerunitbox{\Rbr}
\particleadjusty=-370
\ifodd\unitboxnumber\relax\else\particleadjustx=-240 \particleadjusty=-400 \fi
\or 
\gdef\upperunitbox{\Rbr}   \gdef\lowerunitbox{\Lbr}
\setcoords(-325,-555,0)(-900,-2100,0)[0,-2400]
\particleadjusty=-40
\or 
\setcoords(-505,-275,0)(-900,-2100,0)[0,-2400]
\gdef\upperunitbox{\Lbr}   \gdef\lowerunitbox{\Rbr}
\particleadjusty=-30  
\else \UNIMPERROR
\fi
}
\gdef\SWphoton{  
\ifcase\LINECONFIGURATION  
\setcoords(-825,-825,0)(0,-1250,0)[-1250,-1250]     \SETUNITBOX(1250)[br][tl]  
\or 
\setcoords(-175,-1425,0)(-625,-625,0)[-1250,-1250]  \SETUNITBOX(1250)[tl][br]  
\or 
\setcoords(-900,-900,0)(0,-1410,0)[-1400,-1400]     \SETUNITBOX(1400)[br][tl]  
\or 
\setcoords(-200,-1600,0)(-700,-700,0)[-1400,-1400]  \SETUNITBOX(1400)[tl][br]  
\or 
\setcoords(-800,-800,0)(0,-1200,0)[-1200,-1200]     \SETUNITBOX(1200)[br][tl]  
\or 
\setcoords(-200,-1400,0)(-600,-600,0)[-1200,-1200]  \SETUNITBOX(1200)[tl][br]  
\else \UNIMPERROR
\fi
}
\gdef\Wphoton{
\ifcase\LINECONFIGURATION 
\setcoords(-2245,-1245,0)(-150,-400,0)[-2005,0]
\gdef\upperunitbox{\Frown}   \gdef\lowerunitbox{\Smile}
\or 
\setcoords(-2245,-1245,0)(-400,-150,0)[-2005,0]
\gdef\upperunitbox{\Smile}   \gdef\lowerunitbox{\Frown}
\else \UNIMPERROR
\fi
\particleadjustx=57 
\ifnum\numwigglespo=\number\numwiggles \particleadjustx=0  \fi
\numlowerunits=\numwigglespo   \numupperunits=\numwiggles
}
\gdef\NWphoton{  
\ifcase\LINECONFIGURATION  
\setcoords(-200,-1425,0)(625,625,0)[-1250,1250]   \SETUNITBOX(1250)[bl][tr]
\or 
\setcoords(-825,-825,0)(0,1250,0)[-1250,1250]     \SETUNITBOX(1250)[tr][bl]
\ifnum\numwigglespo > \number \numwiggles \particleadjusty=-15 \fi
\or 
\setcoords(-200,-1600,0)(700,700,0)[-1400,1400]   \SETUNITBOX(1400)[bl][tr]
\or 
\setcoords(-900,-900,0)(0,1400,0)[-1400,1400]     \SETUNITBOX(1400)[tr][bl]
\or 
\setcoords(-200,-1400,0)(600,600,0)[-1200,1200]   \SETUNITBOX(1200)[bl][tr]  
\or 
\setcoords(-800,-800,0)(0,1200,0)[-1200,1200]     \SETUNITBOX(1200)[tr][bl]  
\else \UNIMPERROR
\fi
}